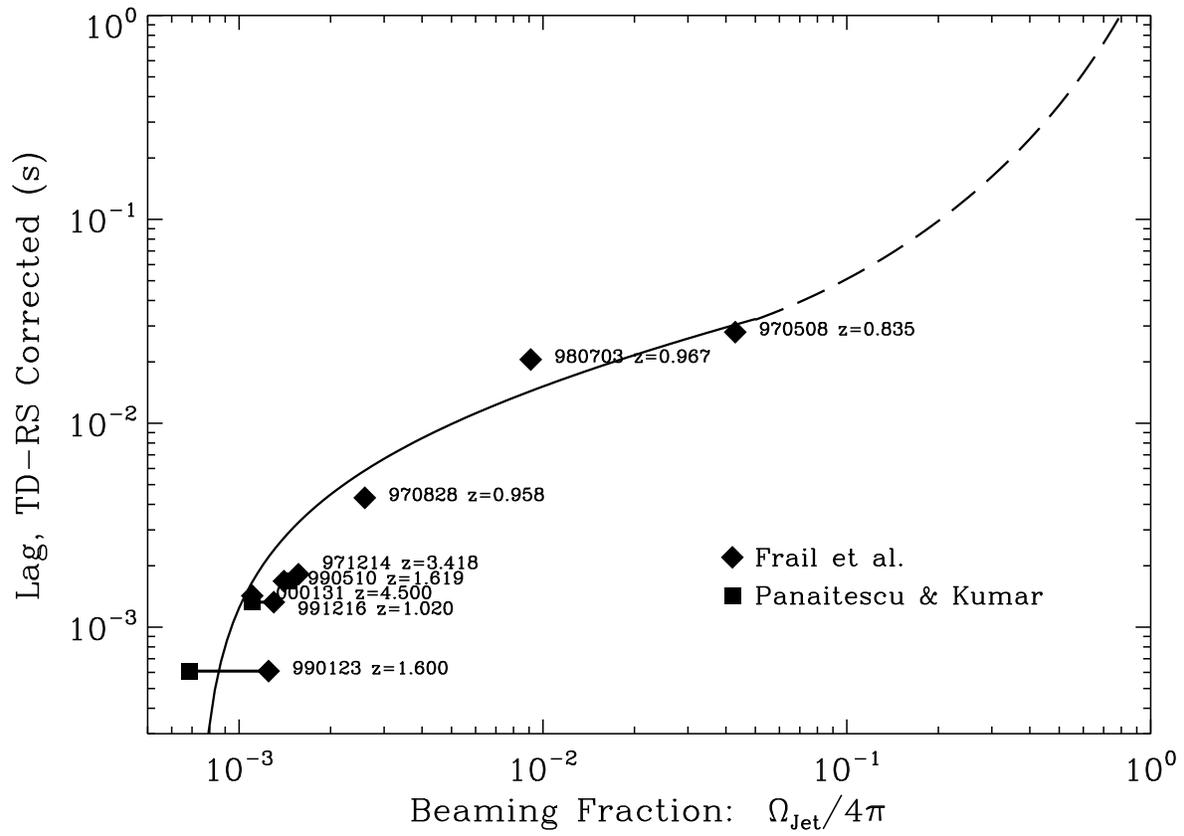

Fig. 1

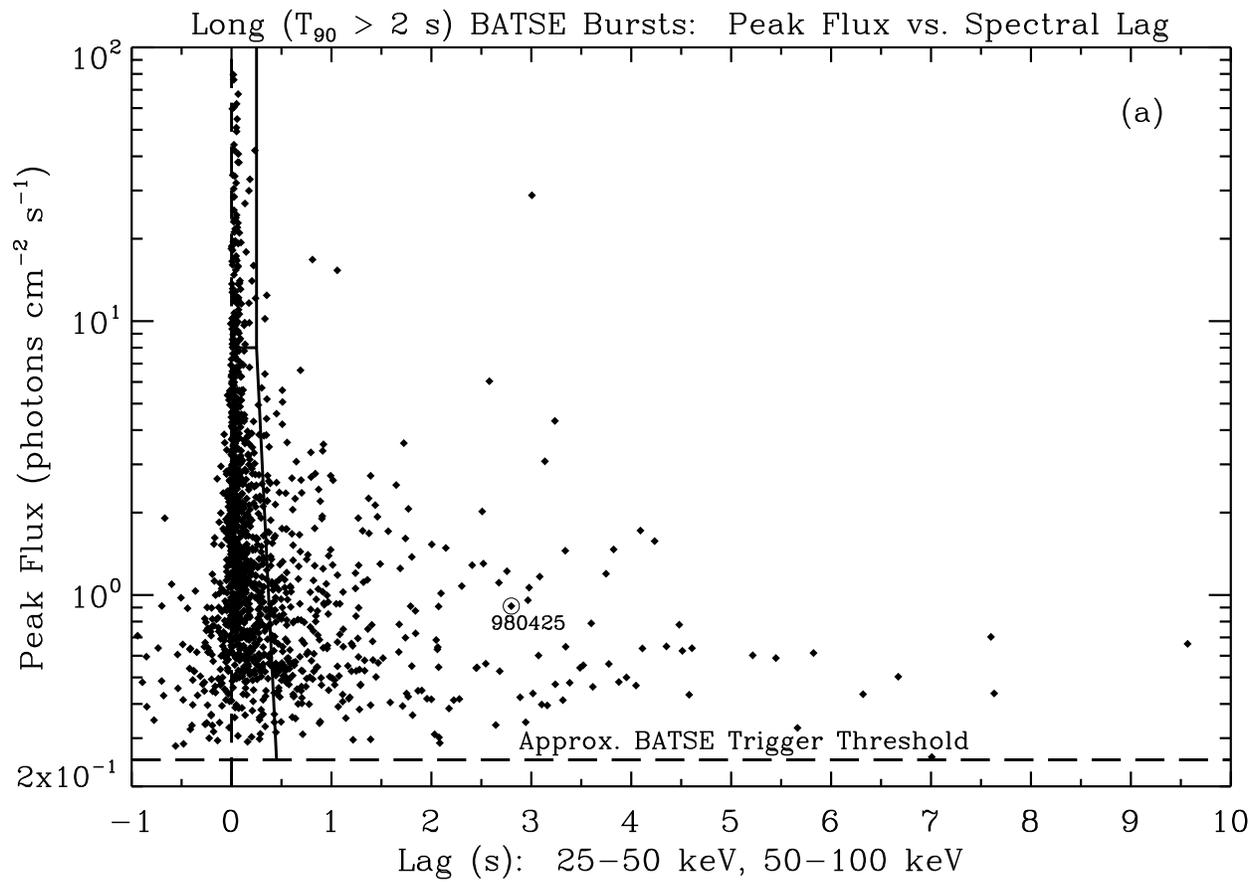

Fig. 2a

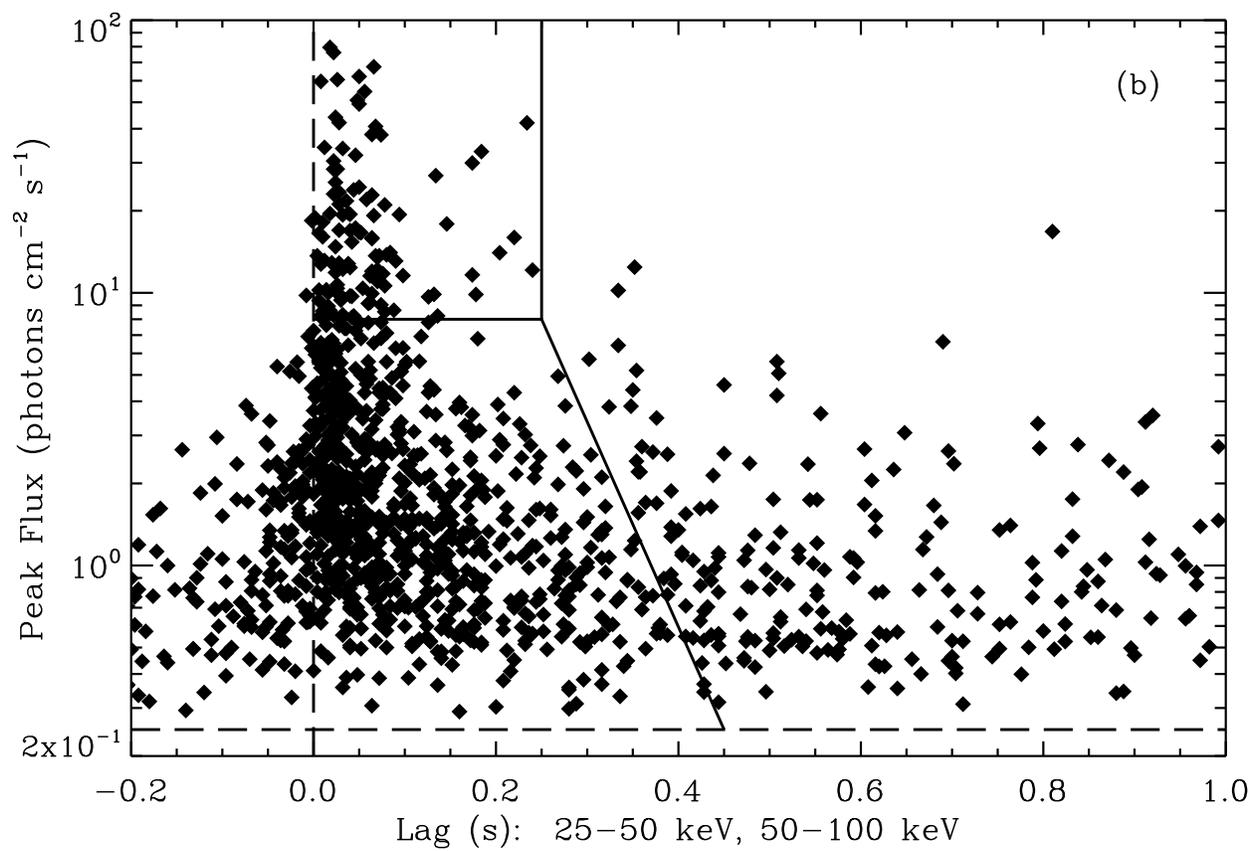

Fig. 2b

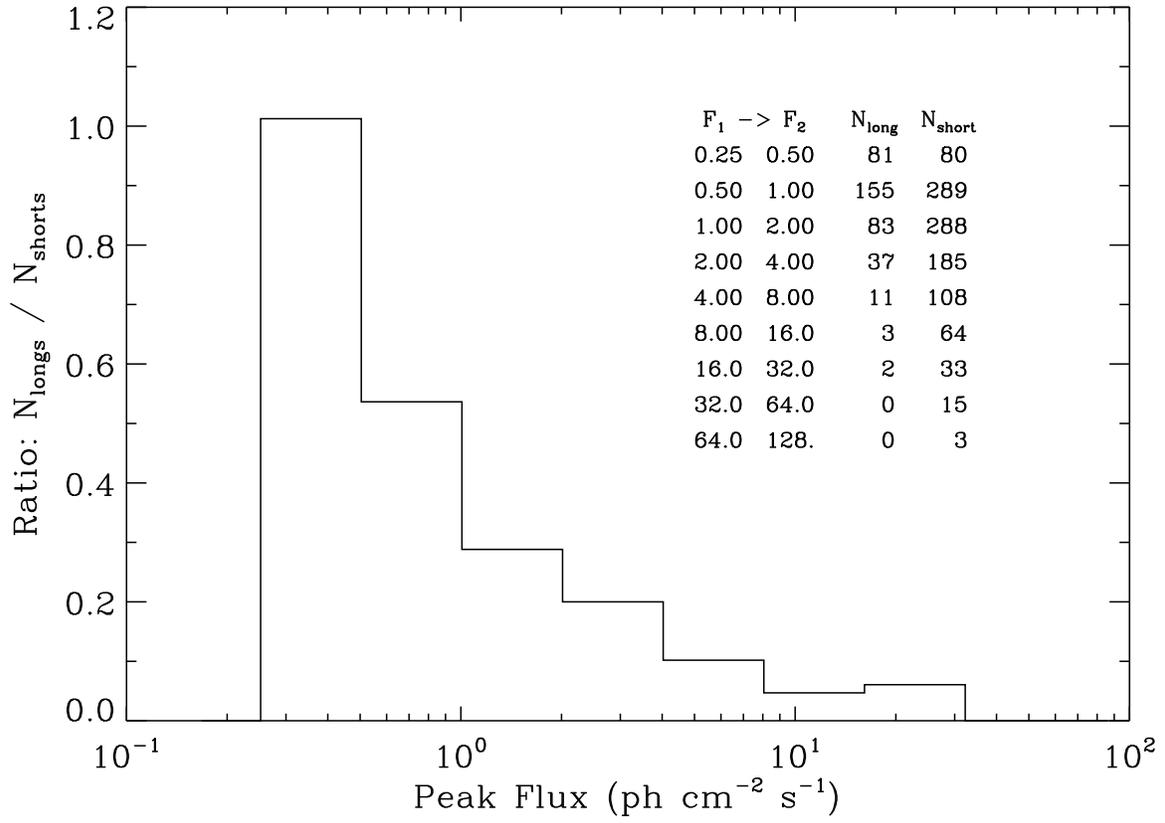

Fig. 3

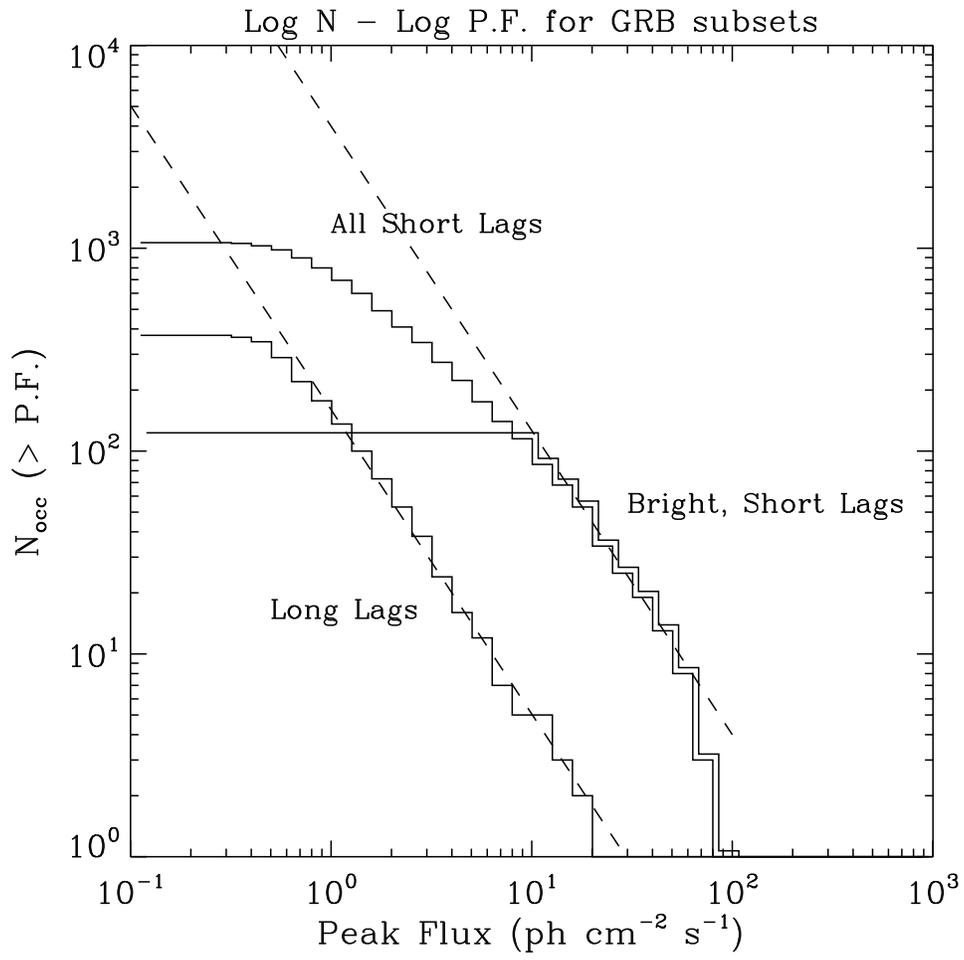

Fig. 4

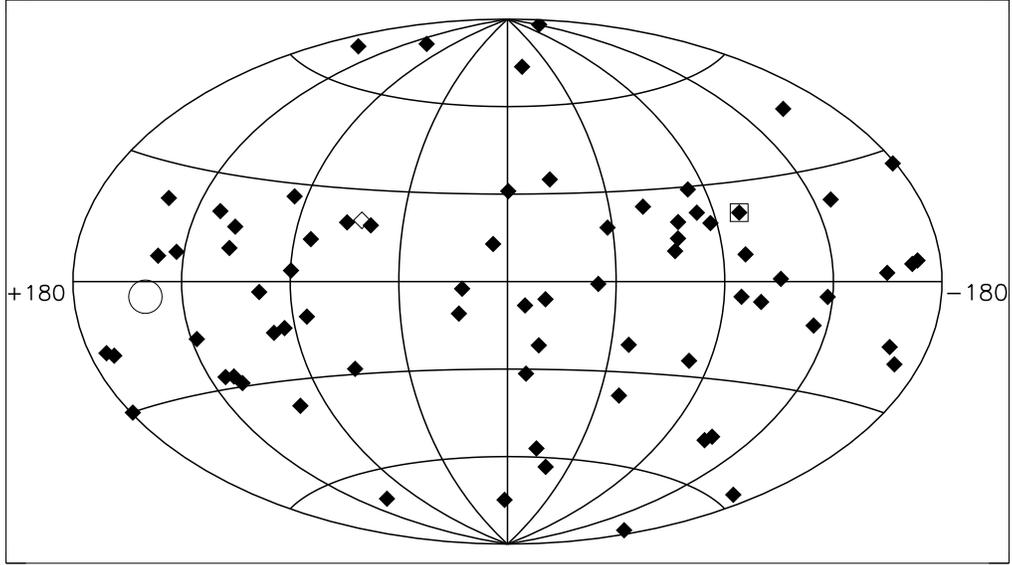

Fig. 5

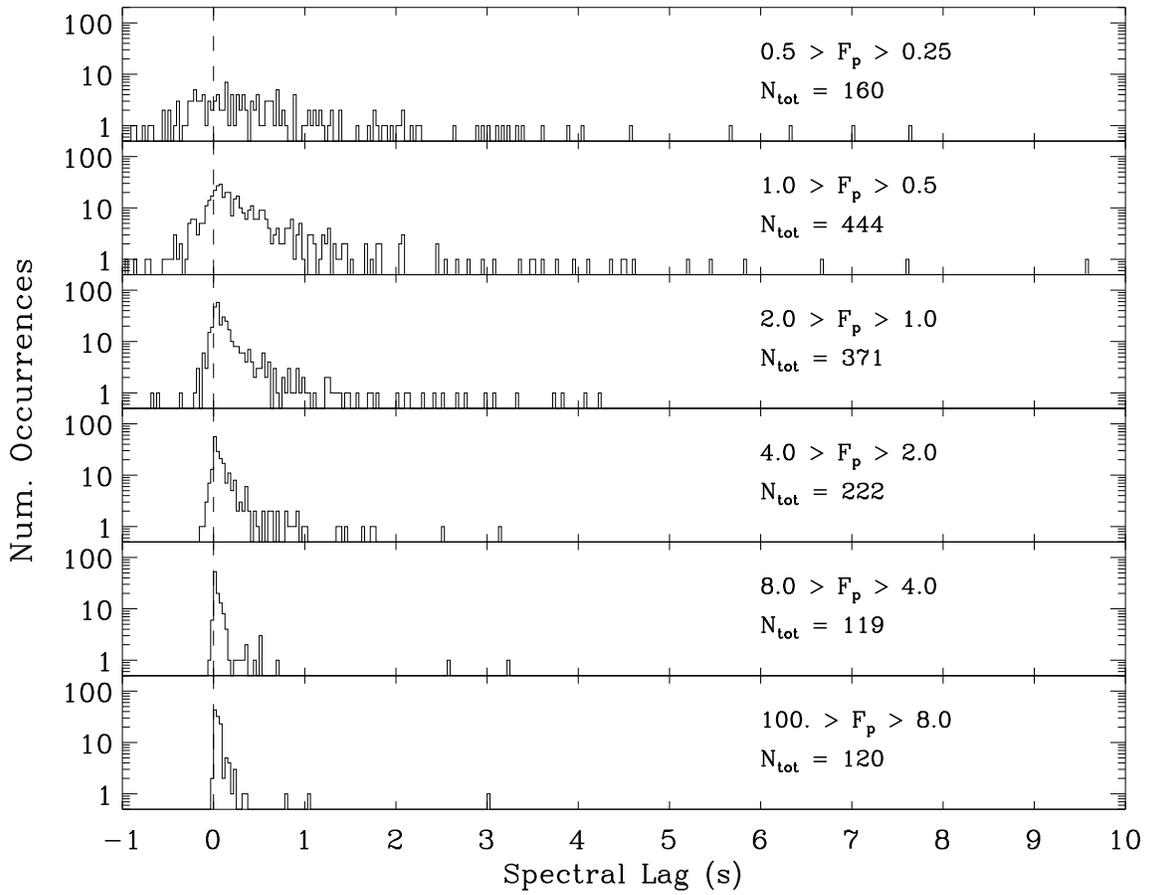

Fig. 6

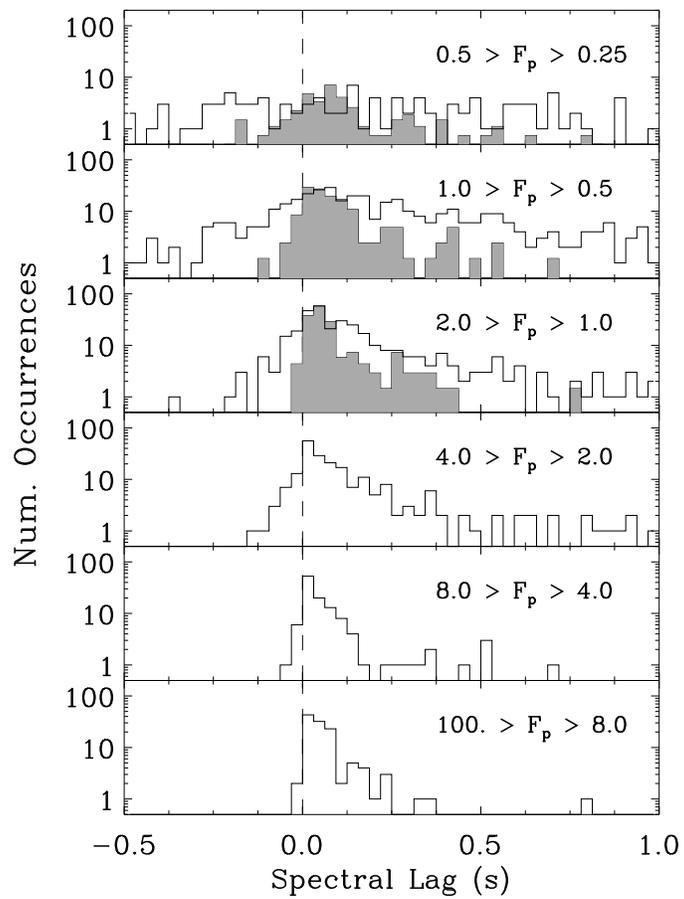

Fig. 7

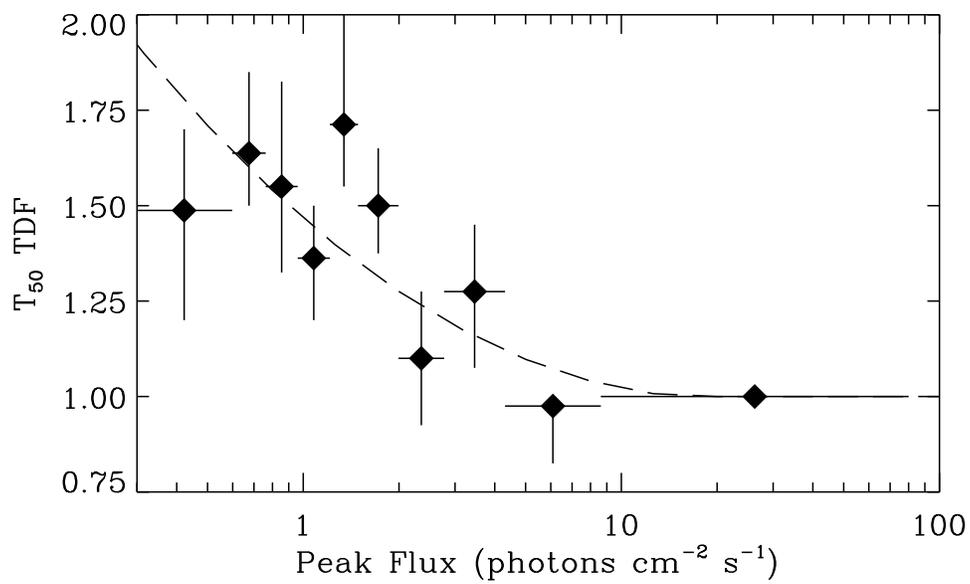

Fig. 8

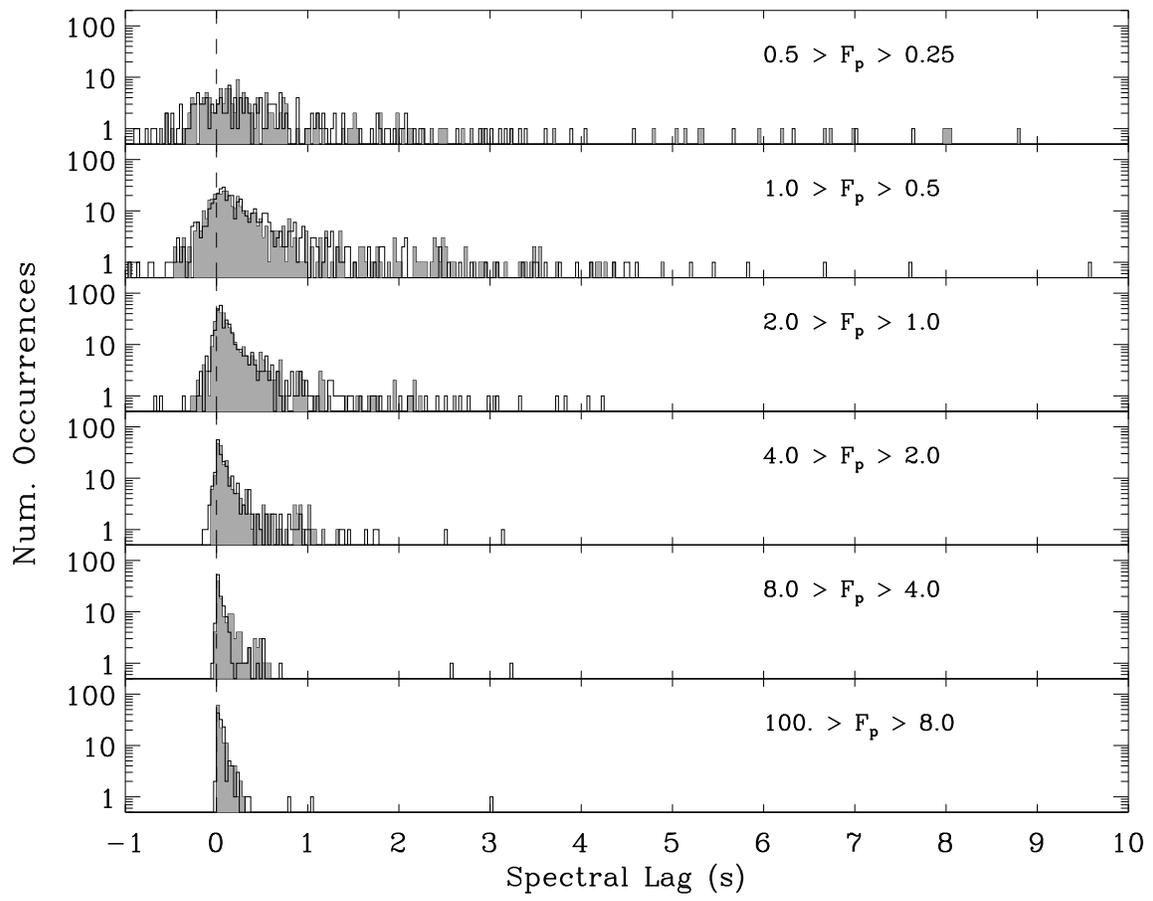

Fig. 9a

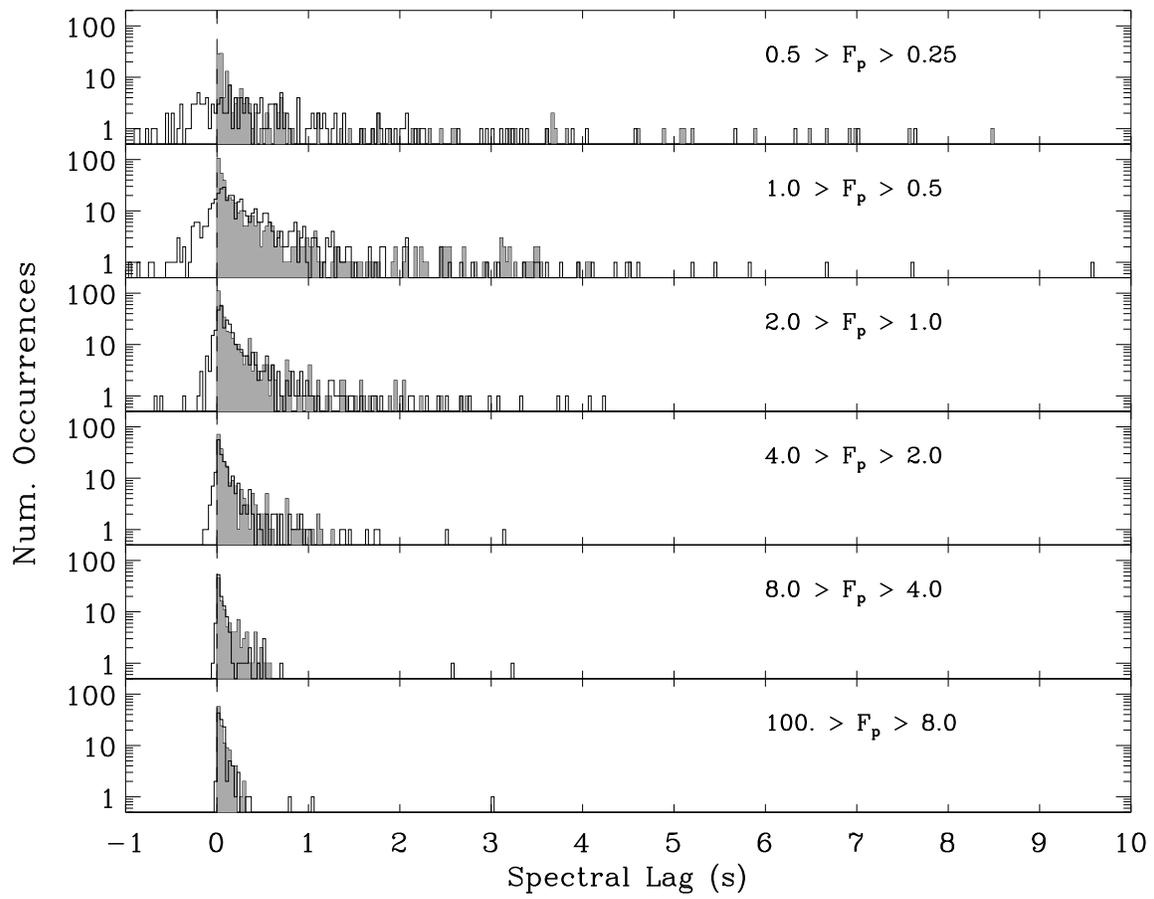

Fig. 9b

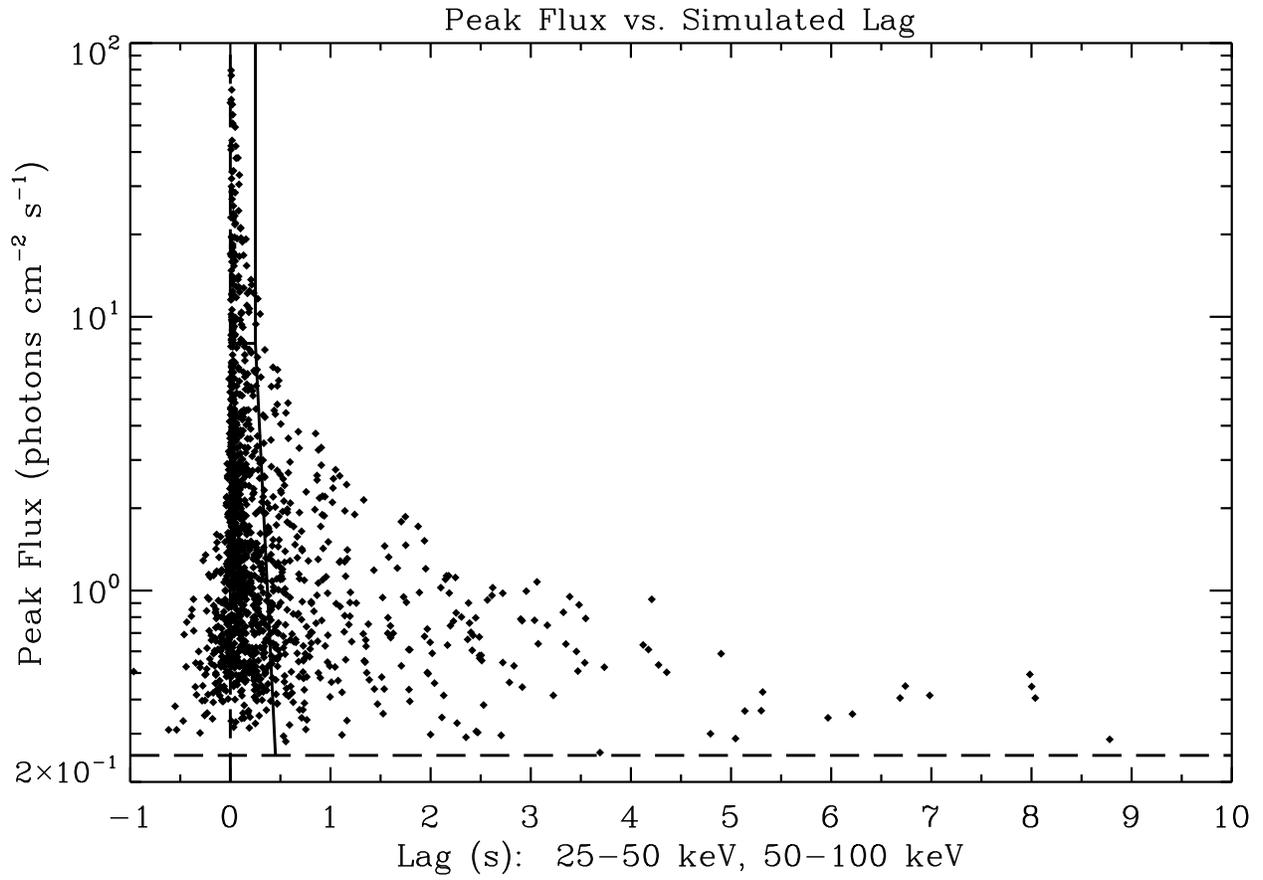

Fig. 10

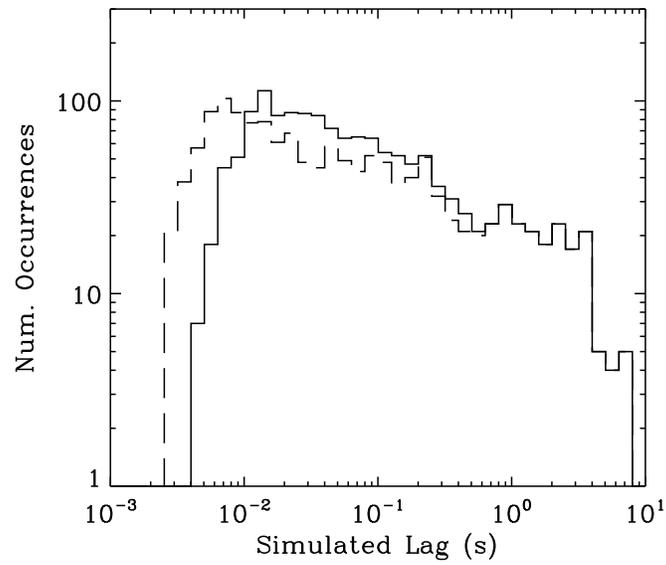

Fig. 11

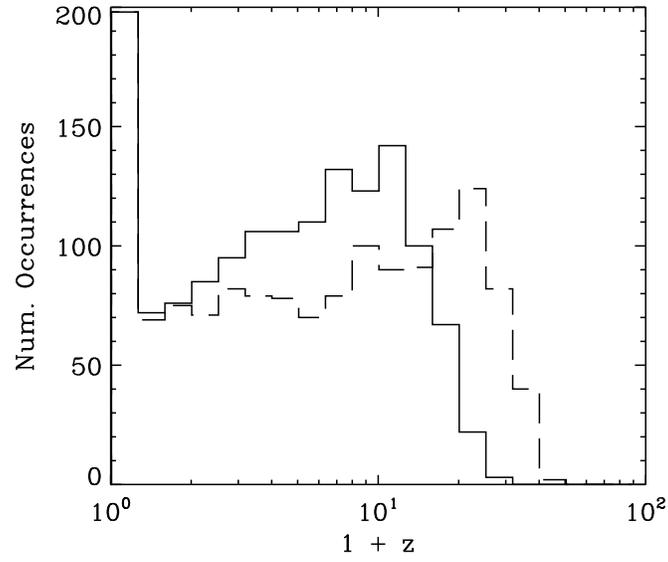

Fig. 12

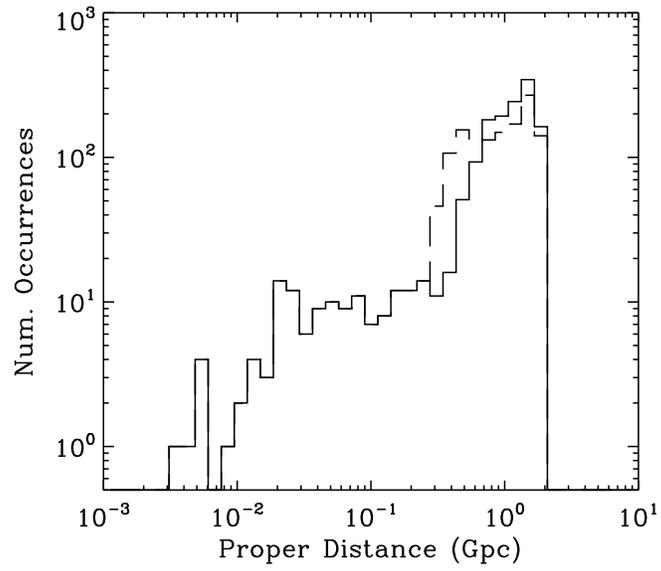

Fig. 13

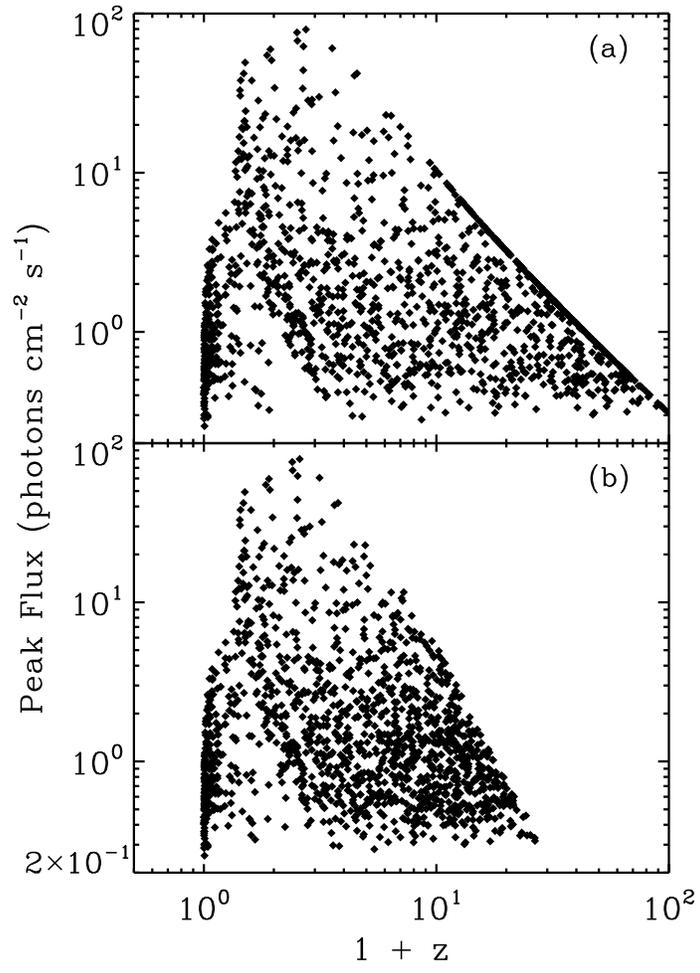

Fig. 14

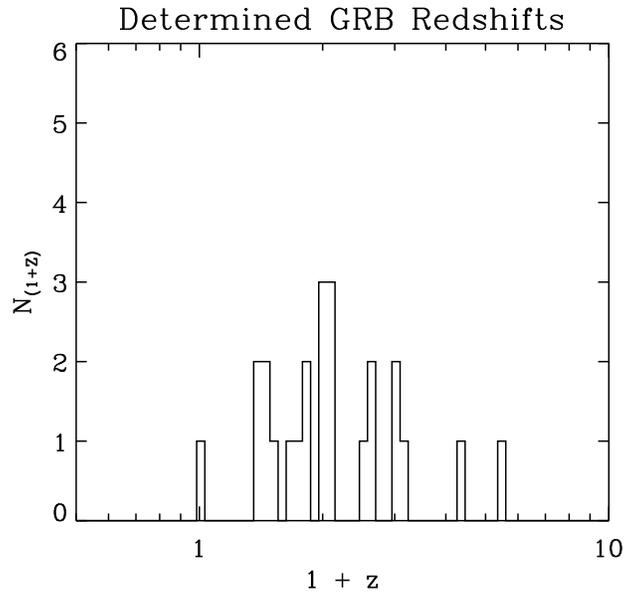

Fig. 15

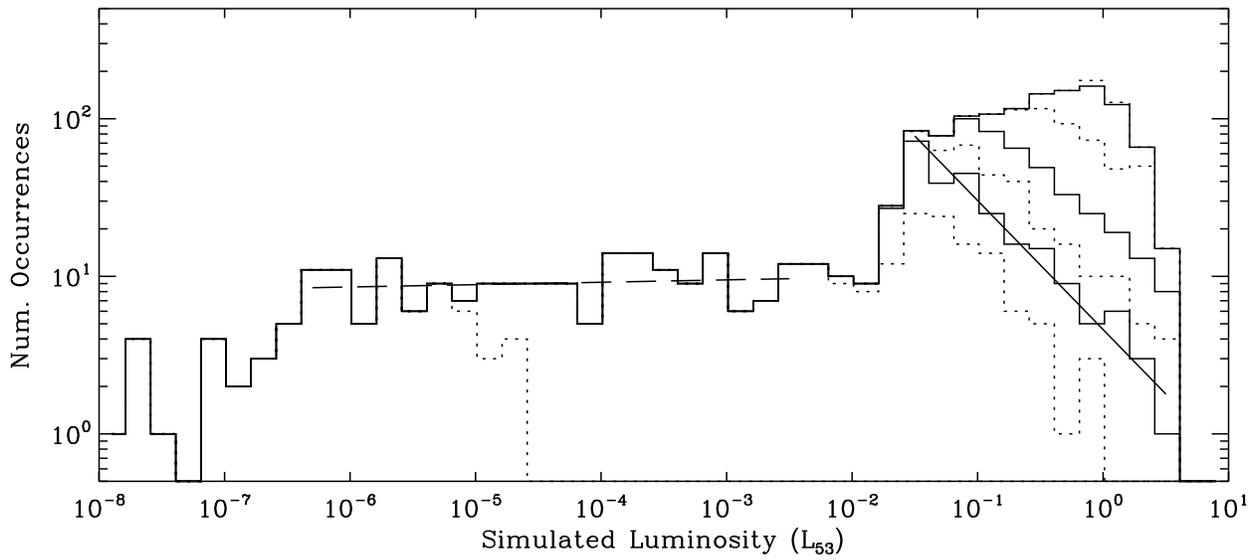

Fig. 16

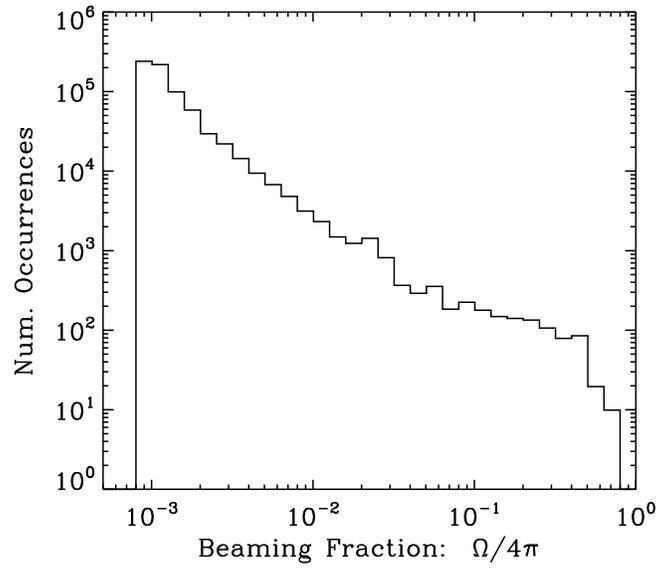

Fig. 17

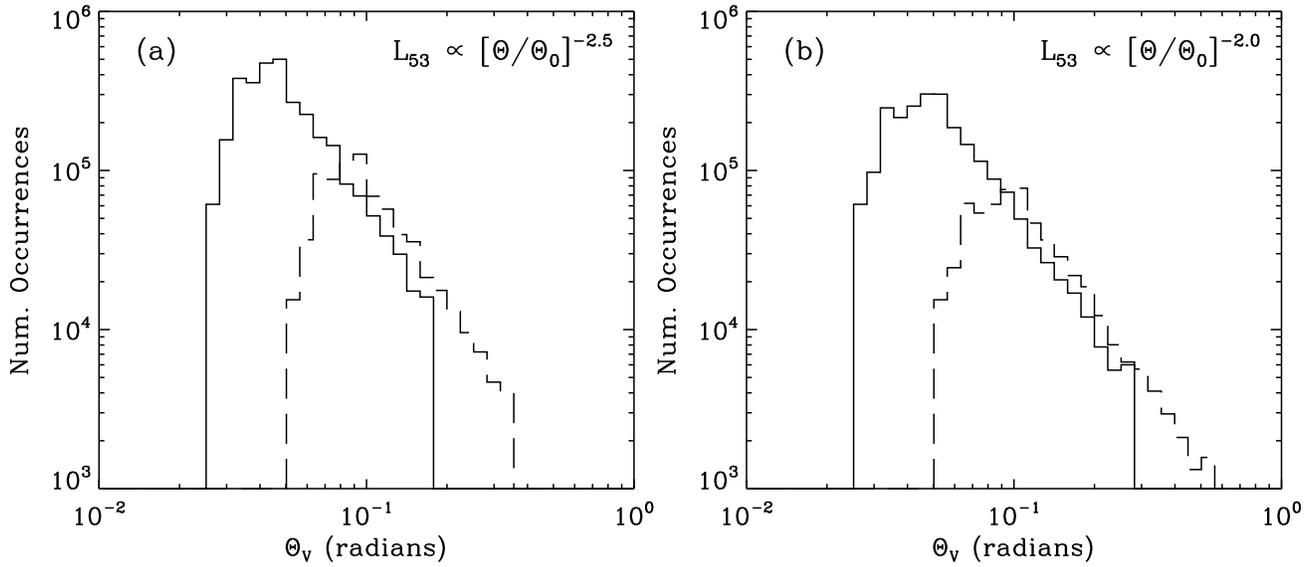

Fig. 18

# IMPLICATIONS OF LAG-LUMINOSITY RELATIONSHIP FOR UNIFIED GRB PARADIGMS

J. P. Norris

[1]NASA/Goddard Space Flight Center, Greenbelt, MD 20771.

ABSTRACT

Spectral lags ($\tau_{lag}$) are deduced for 1437 long ($T_{90}$ > 2 s) BATSE gamma-ray bursts (GRBs) with peak flux $F_p$ > 0.25 photons cm$^{-2}$ s$^{-1}$, near to the BATSE trigger threshold. The lags are modeled to approximate the observed distribution in the $F_p$–$\tau_{lag}$ plane, realizing a noise-free representation. Assuming a two-branch lag-luminosity relationship, the lags are self-consistently corrected for cosmological effects to yield distributions in luminosity, distance, and redshift. The results have several consequences for GRB populations and for unified gamma-ray/afterglow scenarios which would account for afterglow break times and gamma-ray spectral evolution in terms of jet opening angle, viewing angle, or a profiled jet with variable Lorentz factor:

A component of the burst sample is identified – those with few, wide pulses, lags of a few tenths to several seconds, and soft spectra – whose Log[N]–Log[$F_p$] distribution approximates a -3/2 power-law, suggesting homogeneity and thus relatively nearby sources. The proportion of these long-lag bursts increases from negligible among bright BATSE bursts to ~ 50% at trigger threshold. Bursts with very long lags, ~ 1–2 < $\tau_{lag}$ (s) < 10, show a tendency to concentrate near the Supergalactic Plane with a quadrupole moment of ~ –0.10 ± 0.04. GRB 980425 (SN 1998bw) is a member of this subsample of ~ 90 bursts with estimated distances < 100 Mpc. The frequency of the observed ultra-low luminosity bursts is ~ _ that of SNe Ib/c within the same volume. If truly nearby, the core-collapse events associated with these GRBs might produce gravitational radiation detectable by LIGO-II. Such nearby bursts might also help explain flattening of the cosmic ray spectrum at ultra-high energies, as observed by AGASA.

In a regime limited by BATSE sensitivity, $10^{-6.4}$ < $L_{53}$ < $10^{-2.6}$, the model lags predict a power-law scaling relation for the ultra-low luminosity GRBs, $dN_{sen}/dL \sim L^{-1}$, flatter than expected ($dN_{sen}/dL \sim L^{1/6}$) if viewing angle with respect to the jet axis alone governed perceived luminosity. For high-luminosity bursts, in the volume-limited regime z < 2 and $10^{-1.6}$ < $L_{53}$ < $10^{0.6}$, BATSE samples through the distribution and $dN_{vol}/dL \sim L^{-1.8}$, similar to expectations for viewing angle scenarios ($dN_{vol}/dL \sim L^{-2}$). However, in the latter case if the luminosity decreases off axis, $L \sim \theta_{view}^{-\lambda}$ ($\lambda$ > 0), then overproduction of low-luminosity bursts cannot be avoided. Thus, a completely relativistic kinematic explanation for the dynamic range in GRB luminosities is not favored. The variable beaming fraction scenario, with constant luminosity across the jet cone, can fit the high-luminosity bursts with a fairly flat distribution in jet-cone solid angle, $dN(\Omega_{jet})/d\Omega_{jet} \propto \Omega_{jet}^{-0.2}$; for the ultra-low luminosity bursts a distribution that increases is required, $dN(\Omega_{jet})/d\Omega_{jet} \propto \Omega_{jet}^{+0.5}$. Jets with variable luminosity profiles viewed at a range of angles can also reproduce the observed luminosity distributions, such that $L \sim \theta_{view}^{-2.5}$ and $L \sim \theta_{view}^{-1.3}$, for high and ultra-low luminosity regimes, respectively. For both the beaming fraction and profiled jet scenarios, a large fraction of the SN Ib/c population in the Universe would be required to produce the GRBs at cosmological distances, whose rate is estimated to be 1–few $\times 10^6$ yr$^{-1}$. The modeled redshift distribution for GRBs peaks at z ~ 10, with large uncertainty.

Subject headings: gamma rays: bursts, afterglows, beaming — supernovae: type Ib/c — general relativity: gravitational waves — cosmic rays: ultra-high energy

1. INTRODUCTION

Recent theoretical and observational works have appeared which reveal a new understanding of the γ-ray burst (GRB) energy paradigm compared to the original picture developed over the preceding 4–5 years. The new studies conclude that the actual dynamic ranges of energy release and luminosity are relatively narrow, rather than spanning factors of a few hundred as implied if GRB emissions were isotropic. (Much larger inferred dynamic ranges [~ $10^7$] apply if bursts like GRB 980425 should be considered within the same framework.) All the explanations involve anisotropic ejecta. However, they invoke three distinct physical mechanisms, or combinations thereof, and marshal the observational facts to support the differing interpretations. A perceived range in luminosity, and therefore energetics, could result from: variation in jet cone opening angle while maintaining constant total energy release; variation in viewing angle alone; or a profiled jet where the Lorentz factor ($\Gamma$) decreases as viewed off the jet axis. In each scenario, evidence has been cited that unifies the γ-ray spectral-temporal behavior with the afterglow temporal behavior through either pure relativistic kinematics or the jet dynamics. Since each explanation may be viewed as economical – each realizes reduced and narrow ranges in γ-ray energy release and luminosity – observations may provide the key to distinguishing between the possible scenarios. There are > 2700 BATSE triggered bursts but only two dozen GRBs with associated redshifts, and so γ-ray observations may help distinguish between the predictions of the three scenarios vis-à-vis the GRB luminosity distribution as inferred from spectral lags.

The dynamic range in energy for cosmologically distant GRBs is ~ $3 \times 10^{51} - 10^{54}$ ergs, assuming isotropic emission. By analysis of break times in the decays of GRB afterglows, Frail et al. (2001) have inferred a distribution of initial jet opening angles, inversely correlated with total γ-ray energy and luminosity, with most bursts' ejecta exploding into narrow cones, where the smallest half angle $\theta_{jet} > 2$–$3$ . The analysis embodies a crucial assumption – that luminosity (i.e. $\Gamma$) is constant across the jet's cone. The derived beaming fraction, $f_b \equiv \Omega_{jet}/4\pi \approx \_ \theta_{jet}^2$, exhibits a dynamic range commensurate with observed γ-ray luminosities and total energies. The true γ-ray energy release distribution is then inferred by Frail et al. to be much narrower, less than a decade, and centered near ~ $5 \times 10^{50}$ ergs. The temporal signature of afterglow evolution was enunciated by Rhoads (1997; 1999) for constant $\Gamma$ within the jet cone. When the ejecta's Lorentz factor decreases below ~ $\theta_{jet}^{-1}$, manifesting a break in the power-law decay, at that time spherical and jet geometries become distinguishable; transverse expansion of the jet adds to the steepening decay. Panaitescu & Kumar (2001: PK) perform an analysis similar to Frail et al. The three afterglows in common to the two treatments – for which there are also BATSE data – are on the high end of the GRB luminosity distribution, and the derived sets of opening angles do not differ markedly. Figure 1 illustrates derived beaming fraction and spectral lag for eight bursts with associated redshifts. The two quantities appear correlated over a dynamic range of ~ 50 in both coordinates, as would be expected since spectral lag and beaming fraction are inversely correlated with observed luminosity. So even though the distance scales from the source (Piran 2001) for the two phases are disparate in the internal+external shocks paradigm (~ $10^{13-14}$ cm compared to > $10^{16}$ cm), the correlation evident in Figure 1 implies that the γ-radiation dynamics are related to jet dynamics. An adequate parametric fit of the form

$$\tau_{lag} = -0.127 + 1.11 f_b^{3/10}, \quad \text{or}$$
$$f_b = \{[\tau_{lag} + 0.127] / 1.11\}^{10/3} \quad f_b \leq 0.05 \quad (1a)$$

was found for the six bursts with $0.835 \leq z \leq 1.619$, using the geometric means of beaming fraction for the three bursts in common to Frail et al. and PK. GRBs 000131 and 971214 were not included in the fit since redshift corrections to their spectral lag measurements could prove to be substantial (Morales et al. 2002). The redshift correction factors applied to the lags are small (the correction procedure is discussed in section 2.4). The $\tau_{lag}$–$f_b$ fit is indicated by the solid line. The fit nearly asymptotes near $f_b \approx 7.5 \times 10^{-4}$, equivalent to $\theta_{jet} \approx 2.3$, necessarily consistent with the minimum jet cone radii of 2–3 which PK and Frail et al. discuss.

The distribution of BATSE GRB lags described in the next section extends up to 10 s. This is a factor of 50 longer than the longest lag (GRB 970508: 0.2 s, uncorrected for cosmology) plotted in Figure 1. Thus, *if there were one continuous relationship between lag and beaming fraction for GRBs,* one would need to construct a continuation of Eq. (1a) that asymptotes to $f_b$ ~ 1 at lags _ 10 s. The toy model plotted with a dashed line in Figure 1 achieves the continuity and limit requirements:

$$\tau_{lag} = -0.84 + \exp(3. f_b), \quad \text{or}$$
$$f_b = \ln[\tau_{lag} + 0.84] / 3. \quad f_b \geq 0.05. \quad (1b)$$

Such a change in functional form might reflect a modification of the single power-law form of the lag-luminosity relationship in Norris, Marani, & Bonnell (2000: NMB), i.e., to accommodate the ultra-low luminosity GRB 980425 – currently an example of one. Salmonson (2001) argues for a separate lag-luminosity branch for bursts with low Lorentz factors. However, the proposals by Nakamura (1999) and Salmonson (2001) for GRB 980425 involve a large off-jet-axis viewing angle, rather than a large jet opening angle.

Several other investigators have recently discussed alternative interpretations of the afterglow breaks that do not necessarily require a range in $\theta_{jet}$. One alternative to the variable beaming fraction scenario, invoking viewing angle, has been advanced by several authors, mostly in the context of the low luminosity of GRB 980425 (Wang & Wheeler 1998; Höflich, Wheeler, & Wang 1999; Nakamura 1999; Woosley, Eastman, & Schmidt 1999; Salmonson 2001) and extended to explain the lag-luminosity relationship in general (Salmonson 2000; Ioka & Nakamura 2001; Salmonson & Galama 2001). In essence, softer-spectrum, lower-luminosity, longer-lag bursts will be observed as the observer's viewing angle off the jet axis increases, merely due to relativistic kinematics.

The third possibility – a profiled jet with $\Gamma$ decreasing off axis – is also naturally expected from simulations (see e.g., MacFadyen, Woosley, & Heger 2001). Salmonson (2000) and Salmonson & Galama (2001) discuss this scenario as well, and argue that the observer's viewing angle could explain the apparent dynamic range in luminosity, pulse spectral evolution, and afterglow break times. The profiled jet model is developed in more detail in Rossi, Lazziti, & Rees (2002), and discussed by Zhang & Meszaros (2002) in terms of predicted dependence of luminosity on angle with respect to the jet axis.

Independent evidence supports the picture of a small dynamic range in GRB energies. Piran et al. (2001) have shown from analyses of X-ray afterglows that inferred *total* release energies span less than a order of magnitude – irrespective of assumptions concerning magnetic field strength, electron energy distribution, and external medium. Hence, it is necessary to consider carefully the three explanations for a small range in γ-ray energy release and luminosity.

So far, theoretical interpretations of burst pulses reproduce many, but not all of the observed pulse behavior. The empirical schema to be elucidated is that at γ-ray energies, bursts consist of pulses organized in time and energy (Norris et al. 1996). The rise-to-decay ratio is unity or less; as this ratio decreases, pulses tend to be wider, the pulse centroid is shifted to later times at lower energies, and pulses tend to be spectrally softer. However, since ~ 80% of pulses overlap in bursts, an extensive parameter – such as spectral lag – must serve as a surrogate measurement for the spectral dependence of average pulse shape within a given burst. Variability parameters correlated with GRB luminosity have also been reported (Fenimore & Ramirez-Ruiz 1999; Reichart et al. 2000). (Spectral lag, variability, and pulse shape may evolve during a burst, but the degree of evolution is yet to be well quantified for determination of trends in a large sample.) The canonical theoretical schema (e.g., Piran 2001) is that these pulses of γ-ray emission are produced by the "internal" shocks of colliding pairs of relativistic shells ejected by the central engine. The major timescales are accounted for in this picture: overall burst durations and intervals between pulses are related to the activity of the central engine, and pulse widths are related to the shell collision timescale (Rees & Meszaros 1994; Ramirez-Ruiz & Merloni 2001). Details of pulse shape theory are incomplete. A shift in pulse *peak* as a function of energy is not produced by pure relativistic kinematics in the standard picture where the observer is within the jet's half angle, $\theta_{jet}$ (Fenimore, Madras, & Nayakshin 1996; see also, modeling by Panaitescu &

Meszaros 1998). Outside of this cone, the observed pulse is broadened as viewing angle increases, and peak shift may be effected by assuming an inhomogeneous Lorentz factor over the face of the jet, but this "beyond the jet cone" emission quickly falls below the γ-ray portion of the spectrum (see Nakamura 1999; Ioka & Nakamura 2001; Salmonson 2001; Salmonson & Galama 2001; and Kumar & Panaitescu 2000 for off-axis afterglows). Also, the observed locus in time of pulse intensity versus peak in ν F(ν) – essentially defining pulse shape as function of energy and time – is not reproducible by relativistic kinematics, even when combined with variations in Γ or ν F(ν) across the blastwave emitting surface and with variable shell thicknesses; the implication is that pulse spectral evolution requires attention to in situ cooling details (Soderberg & Fenimore 2001). And, while spherical blastfront curvature relates the Lorentz factor to the *observable* emission cone ($\theta_\gamma \sim 1/\Gamma$) by the relativistic trigonometry – yielding a simplistic, energy-independent form for pulse width, $\Delta t_{pulse} \sim R_{shock}/2c\Gamma^2$ – only recent discussions have connected jet dynamics to luminosity, pulse spectral lag and/or variability (Zhang & Woosley 2002; Rossi et al. 2002).

It has been suspected that a subclass of ultra-low luminosity, very long-lag, soft-spectrum, nearby GRBs exists – GRB 980425 being the well-known exemplar – that could be produced by a version of the collapsar model (MacFadyen, Woosley, & Heger 2001). Beyond the deduction that such bursts should have low Lorentz factors (Γ ~ few: Kulkarni et al. 1998; Woosley & MacFadyen 1999; Salmonson 2001) compared to $\Gamma \sim 10^2$–$10^3$ for the high-luminosity bursts at cosmological distances, additional explanations advanced for low observed luminosity include viewing angle, profiled jets, or much wider jet opening angles. Kulkarni et al. and Wieringa, Kulkarni, & Frail (1999) infer the latter for GRB 980425, based on the conclusion that the radio emission was not strongly beamed. Thus GRBs with very disparate observed luminosities, ranging over ~ $10^{47}$–$10^{53}$ ergs s$^{-1}$, may manifest a range of jet opening angles, or they may produce profiled jets with variable Γ which are viewed from different angles. The whole GRB population might be unified in one of these senses.

Also, it is clear that pulse frequency of occurrence within a given burst depends on peak flux. Norris, Scargle, & Bonnell (2001) showed in a brightness-independent analysis (to peak flux $F_p >$ 1.3 photons cm$^{-2}$ s$^{-1}$) that BATSE bursts with relatively long lags tend to have just a few significant, wide pulses and that such bursts are observed preferentially at lower peak fluxes. This result was presaged by the burst "complexity parameter" of Stern, Poutanen, & Svensson (1999), whose analysis of average GRB profiles as a function of peak flux suggested an admixture of a larger fraction of simple bursts near the BATSE trigger threshold, $F_p \sim 0.25$ photon cm$^{-2}$ s$^{-1}$. A hint that these simple, dim bursts come from sources at low redshifts is provided by the only exemplar with known distance, GRB 980425 / SN 1998bw. Its ultra-low luminosity may be attributable to membership on a steeper branch in the lag-luminosity "HR diagram" of GRBs (see

Salmonson 2001; also Kulkarni et al. 1998). The steeper slope of this second branch may be related to mildly relativistic outflow; if the initial ejecta have $\Gamma^{-1} > \theta_{jet}$, then different behavior is expected for the γ-ray and immediate afterglow phases. Central questions concerning these bursts with relatively simple temporal profiles include their observed frequency, typical beaming fraction and luminosity, and the implied volume sampled by BATSE.

The program here is to estimate luminosities and distances from measured spectral lags for long ($T_{90} > 2$ s) BATSE bursts, and derive constraints on the GRB population and jet mechanism. In Norris, Marani, & Bonnell (2000: NMB) an anti-correlation between γ-ray spectral lag and GRB peak luminosity was reported, based on six bursts with associated redshifts. The relationship is roughly determined as $L_{53} \Box \tau_{lag}^{-1.15}$. The trend is qualitatively strengthened by the addition of two bursts: GRB 991216, $\tau_{lag} \sim 10$ ms, $L_{peak} \sim 6 \times 10^{52}$ ergs s$^{-1}$; and GRB 000131, $\tau_{lag} \sim 5$ ms, $L_{peak} \sim 10^{54}$ ergs s$^{-1}$. The original lag values (and the two listed here) reported for this relationship were corrected for time dilation, but not for spectral redshift. Direct methods for spectral correction involve attempts to deredshift the spectra and measure lags, or to construct interpolation tables with lag measured between several pairs of energies (Morales, Norris, & Bonnell 2000). Here, only bursts within a range of $(1+z_{max})/(1+z_{min}) = 1.4$ are used to define a relationship between lag and beaming fraction, since shifts of the spectral energy distribution for GRB 971214 ($z = 3.14$) and GRB 000131 ($z = 4.5$) are large compared to the median redshift ($z \sim 1$) for the balance of the sample. An indirect method for approximating spectral redshift correction used the average width of the auto-correlation function versus energy, $W_{<ACF>} \Box E^{-0.40}$ (Fenimore et al. 1995). Similarly, the average pulse width as a function of energy can be utilized, $W_{<pulse>} \Box E^{-0.33}$ (Norris et al. 1996), as is done here. Both functions were derived using temporal profiles of bright bursts and, strictly speaking, are applicable over a relatively narrow range in redshift. Thus, calibration of the lag-luminosity relation and necessary spectral corrections will benefit from denser observations in frequency, over a broader dynamic range, and from a larger burst sample with redshifts.

Section 2 describes the extension of spectral lag analysis for BATSE bursts to near the instrument's trigger threshold. The lags are modeled, including noise, to reproduce the observed scatter plot in the $\tau_{lag}$–$F_p$ plane. The Log[N]–Log[$F_p$] distribution and sky distribution of the long-lag burst subsample are examined. Then assuming a two-branch lag-luminosity relation, approximate correction is made for cosmological effects, and the model lags are used to estimate distributions of luminosity, redshift, distance, and beaming fraction. In section 3 the modeled luminosity distribution is compared with expectations for the three jet scenarios described above. In section 4 the results are summarized. Some topics in high-energy astrophysics related to GRB studies are briefly discussed, including: the GRB–SN Ib/c connection; rates for observable gravitational radiation from GRB within the collapsar scenario; the Swift GRB yield; and finally the possibility of ultra-high energy cosmic rays from nearby GRBs.

## 2. SPECTRAL LAG ANALYSIS

The first step is to measure spectral lags for nearly the complete sample of BATSE triggered GRBs with $T_{90}$ durations > 2 s. Then using a small number of parameters, the lag–peak flux distribution is modeled, in the process realizing a noise-free representation for the lags. An iterative procedure is used to correct these modeled lags for the extrinsic cosmological effects, while simultaneously unfolding approximations for the GRB luminosity and redshift distributions. This procedure requires an assumption about the correction required for redshift of burst spectra, and so yields parameter-dependent distributions, rather than unique results.

Beyond selecting bursts with $T_{90}$ > 2 s, the sample is further restricted by requiring $F_p$ (50–300 keV) > 0.25 photons cm$^{-2}$ s$^{-1}$ (measured on 256-ms timescale), and peak intensity (PI) > 1000 count s$^{-1}$ (> 25 keV). Background fits and burst regions were defined, and peak fluxes and durations measured following the same procedures described in Norris et al. (1996) and Bonnell et al. (1997). Starting with an available sample of 2699 BATSE bursts, 2024 survived with usable, concatenated DISCLA, PREB, and DISCSC data, and with satisfactory background fits; 1474 of these were measured to have $T_{90}$ > 2 s. Twenty bursts had 700 < PI (count s$^{-1}$) < 1000; within this lowest count rate range the lag analysis becomes increasingly less useful.

### 2.1 *Cross-Correlation Lag Analysis*

A cross-correlation analysis of BATSE channels 1 (25–50 keV) and 3 (100–300 keV) was performed, as described in detail in NMB. The peak in the cross-correlation function (CCF) was taken as the measure of spectral lag. Only three important modifications to the original procedure were implemented. First, the native 64 ms data were binned to 128 ms (256 ms) resolution for bursts with PI below 7000 (1400), to facilitate location of the central peak in the CCF at low intensities. Second, the 101 realizations per burst with added Poisson-distributed noise (per energy channel), used to estimate statistical errors for the CCF, were restarted with a longer fitted range near the peak of the CCF if the fit was concave up. Third, a trick was used to eliminate a problem in the IDL polynomial function which can occur when the independent variable range is too narrow compared to that of the dependent variable: the time coordinate was expanded by a factor of ten, and the fitted polynomial coefficients were adjusted accordingly. The CCF was computed over the portion of a burst temporal profile extending out to the furthest points attaining half of the peak intensity. Utilizing outlying, lower intensity portions of a burst results in larger lag errors for the majority of bursts treated here. A cubic fit to the CCF was employed to accommodate the asymmetric nature of bursts on all timescales (Nemiroff et al. 1994). Seven bursts were eliminated when, during the lag measurement process, the data were found to be corrupted by either very intense intervals (and therefore counter overflow) or by electronic glitches induced in the DISCSC data by gaps in other data types. The discovery in each case resulted from investigating an apparently significant (but spurious) negative lag. In ten cases the CCF did not consistently fit a peak in at least 50 of 101 iterations (the program found the "peak" at the edge of the CCF), and therefore the burst was eliminated. The final sample with measured lags contained 1437 long bursts.

Figure 2a illustrates the computed lags versus peak flux in a linear-log plot. In this rendition, the lags (uncorrected for cosmology) for the BATSE bursts with redshifts (see Figure 1) span a relatively narrow zone – but still a factor of ~ 40 (0.01–0.37 s). Yet the longest lag for a burst source at high redshift is still 25 times shorter than the longest lag in the sample, ~ 10 s. But, relatively short lags dominate the BATSE sample (see Band 1997): more than 1000 bursts have $\tau_{lag}$ < 350 ms. The $\tau_{lag}$–$F_p$ plane is shown divided by solid lines into three regions containing: 120 bright bursts with $F_p$ > 8 photons cm$^{-2}$ s$^{-1}$ and $\tau_{lag}$ < 0.25 s, whose error bars are comparable to or smaller than the symbols in Figure 2a; 945 dimmer bursts with the maximum $\tau_{lag}$ ranging from 0.25 s to 0.45 s (as $F_p$ decreases from 8 to 0.25 photons cm$^{-2}$ s$^{-1}$), where the error bars increase significantly, with some extending to ~ –1 s; and 372 bursts with longer lags, and error bars typically $\Delta\tau_{lag}/\tau_{lag}$ ~ 25%. The point for GRB 980425 / SN 1998bw is circled ($\tau_{lag}$ = 2.8 s, $F_p$ = 0.9 photons cm$^{-2}$ s$^{-1}$). Figure 2b with the lag coordinate magnified shows more clearly the dense region of low peak flux and short lag. The dispersion in lag towards negative values is attributable completely to measurement error, as demonstrated in section 2.3.

## 2.2 *Long-lag Bursts*

The solid line with negative slope in Figure 2b (dividing the second and third regions of the $\tau_{lag}$–$F_p$ plane) was positioned to take into account the larger lag errors at lower peak fluxes. This is an attempt to separate the "long lag" bursts, which begin to dominate at low peak flux, from the dim, short lag bursts. Notice that below $F_p$ ~ 0.6 photons cm$^{-2}$ s$^{-1}$ the frequency of bursts of all lags clearly diminishes. This reflects the decreasing completeness of the BATSE sample as the trigger threshold is approached. More meaningful is the ratio of number of long-lag bursts (region 3) to short-lag bursts (regions 1+2), shown in Figure 3 as a function of $F_p$ in dyadic steps. This ratio increases dramatically from zero for the brightest bursts, to unity near trigger threshold. Figure 4 illustrates the integral peak flux–frequency distributions (Log[N>$F_p$]–Log[$F_p$]) for the long-lag bursts, short-lag bursts, and bright short-lag bursts. The long-lag bursts follow a –3/2 power-law over ~ 1_ decades in $F_p$ (to much lower in $F_p$ than do the bright bursts), with the inevitable rollover consistent with trigger threshold effects. The implications are that a relatively nearby GRB component of low luminosity is being detected, and that this component begins to dominate the frequency of burst detection near BATSE threshold. These long-lag bursts have the softest spectra of all three regions delineated in Figure 2a (Bonnell & Norris 2002).

GRB 980425 is the canonical long-lag, soft-spectrum, ultra-low luminosity burst, its source lying only ~ 38 Mpc distant (Galama et al. 1998). Therefore, it is necessary to examine the sky distribution of the long-lag bursts in Supergalactic (SG) coordinates (Hudson 1993). Table 1 lists the quadrupole moment (Q) and error ($\varepsilon_Q$) for long-lag bursts for $\tau_{lag}$ > 0.5–3.0 s in 0.5-s steps, along with number of bursts ($N_{GRBs}$) and formal significance expressed in standard deviations (see Hartmann et al. 1996). The quadrupole moment for the 1065 bursts with short lags (those bursts in regions 1+2 discussed above) is $Q_{short\_lag}$ = -0.005 ± 0.009. Since these bursts are believed to be at cosmological distances, $Q_{short\_lag}$ constitutes an empirical measure of the quadrupole moment for BATSE exposure. In Table 1 the Q values were reduced by $Q_{short\_lag}$, and the exposure error was propagated in quadrature to obtain the $\varepsilon_Q$ values. For very long-lag bursts, $\tau_{lag}$ > 1.5 s, the significance of the quadrupole moment is as high as 2–2.7 σ before dropping as the number of bursts becomes small. Figure 5 shows the SG distribution of the 72 bursts with $\tau_{lag}$ > 2 s; the concentration of this sample towards the SG plane is evident, with _ of the sources occupying that half of the sky between –30 and +30 in supergalactic latitude. The center of the Virgo Cluster is indicated by the large open circle. GRB 980425 / SN 1998bw lies at ($b_{SG}$, $l_{SG}$) = (20.8, -100.9).

Taken together, the guilt by association with GRB 980425, the approximately –3/2 power law for all long-lag bursts, the preferential detection of these bursts near BATSE threshold, and the tendency of very long-lag bursts to follow the main feature of the nearby matter distribution – while not incontrovertible evidence – suggest that very long lag implies ultra-low luminosity. The implications of this conclusion are explored in the Discussion. Also indicated in Figure 5 is the position of GRB 971208, probably the BATSE burst with the longest spectral lag (visual estimate ~ 20–30 s) and median peak flux, ~ 1.3 photons cm$^{-2}$ s$^{-1}$. Its mono-pulse temporal profile spanned an Earth occultation (Connaughton et al. 1997), and was so long that it was not included in the lag analysis performed here since a usable background could not be fitted.

### 2.3 *Modeling Spectral Lag*

The next step is to model the $\tau_{lag}$–$F_p$ scatter plot of Figure 2a with few parameters. The motivation is to realize a noise-free representation of the form $N(\tau_{lag}, F_p) = f(F_p)$, which can be used to estimate redshifts and luminosities. Construction of a satisfactory representation of $N(\tau_{lag}, F_p)$ is facilitated by dividing the $\tau_{lag}$–$F_p$ plane into peak flux ranges in dyadic steps, as illustrated in Figure 6. The increase in size of measurement error is evident from the larger dispersion in lag near $\tau_{lag} = 0$ for successively dimmer $F_p$ ranges. However, the dispersion cannot be completely reproduced by assuming that lags for dimmer bursts follow the same distribution as lags for bright bursts ($F_p > 8.$ photons cm$^{-2}$ s$^{-1}$) but with lower S/N. This is demonstrated in Figure 7, where the lags for the 120 brightest bursts have been recomputed for the three lowest $F_p$ ranges (0.25–0.5, 0.5–1.0, and 1.0–2.0 photons cm$^{-2}$ s$^{-1}$) but with peak intensities and S/N levels chosen randomly from bursts in the respective $F_p$ ranges (see Norris et al. 1994 for a description of the Poisson S/N-equalization procedure). The bright-burst lag histograms are gray-filled and their peaks are normalized to the peaks of the solid-line histograms of the dimmer subsamples. In the core, $\tau_{lag} < 0.5$ s, the distributions for S/N-equalized bright bursts are significantly narrower than the distributions for the three dim $F_p$ ranges, indicating that dimmer bursts tend to have a longer lag cutoff in the core component. Presumably, this reflects the detection of lower luminosity bursts at lower $F_p$.

A model for the observed lags must reproduce the increasing frequency of longer lags at lower peak fluxes, both in the core and extending across the tail of the distribution to $\tau_{lag} \sim 10$ s. It is possible to reproduce both aspects using one power-law distribution, but with a long-lag cutoff, $\tau_{max}$, which increases with decreasing $F_p$.

The minimum lag for bright bursts, $\tau_{min0}$, is another fit parameter. This minimum then would correspond to some maximum luminosity. However, just like other spectral-temporal measures – e.g., average peak-aligned profile, duration, pulse interval – for which observed time-dilation trends with peak flux have been observed (Norris et al. 1994; Bonnell et al. 1997; Deng & Schaefer 1998), spectral lag most probably exhibits a similar average stretching trend as peak flux decreases: lag is a surrogate measure for pulse spectral evolution. Wider pulses have longer lags, and pulse width is affected by the twin cosmological effects of time dilation and spectral redshift. At this point in our understanding, it is not clear whether the observed time dilation is extrinsic (due to expansion of the Universe) or intrinsic (due to burst dynamics and/or observer viewing angle). Yet the question is presently irrelevant, since the modeling should take into account the observed trend regardless of cause. The problem is how to represent an observed time dilation given that most spectral lags are short compared to most other timescales in bursts, and that lag measurements at low peak flux have large associated errors. The several measures of extrinsic time dilation/spectral redshift depend (nonlinearly) in different ways on the two effects (Norris 1996), but yield comparable trends with peak flux. One reliable expedient is to use time-dilation factors (TDFs) measured for $T_{50}$ durations; the uncertainties of this method are well understood (e.g., Bonnell et al.). Figure 8 shows the average $T_{50}$ durations for 10 peak flux ranges, each group containing $\sim 100$ bursts. Only the bursts in regions 1 and 2 of Figure 2 were used for the purpose of estimating time dilation; the long-lag bursts were excluded based on the conclusions reached in section 2.2. A quadratic form was fitted to the TDF–$F_p$ trend, and $\tau_{min}$ was modeled as a (fixed) function of peak flux to yield an empirical stretch factor, $S_{obs} = a + b\,F_p + c\,F_p^2$, with $a = 1.47$, $b = -0.722$, and $c = 0.277$.

Five parameters are then required to describe the modeled lag distribution – $\tau_{min0}$, $\tau_{max0}$, $\alpha$, $\beta$, and $F_{p0}$ – such that

$$N(\tau_{model}, F_p) = \tau_{model}^{-(\alpha+1)}$$
$$\text{with } \tau_{min}(F_p) = \tau_{min0}\,S_{obs} \qquad F_p < 25$$
$$\tau_{min}(F_p) = \tau_{min0} \qquad F_p \geq 25$$
$$\text{and } \tau_{max}(F_p) = \tau_{max0}\,(F_p/F_{p0})^\beta \ . \qquad (2)$$

Integration and inversion results in the Monte Carlo formulation for the modeled lag,

$$\tau_{model} = \tau_{min}\,\{\,1 - R\,[1 - (\tau_{max}/\tau_{min})^{-\alpha}]\,\}^{-1/(\alpha+1)} \qquad (3)$$

where R is a uniformly distributed random deviate. Thus, while a single power-law index, $\alpha$, describes the lag distribution, the relative density of short lags thins out as $\tau_{max}(F_p)$ increases.

Several attempts were made to model the lag errors analytically as a function of lag and peak flux; none provided adequate results. Instead, the model errors were gotten from the real error sample: For each model lag, an error for an associated real lag was chosen randomly from a region in the $\tau_{lag}$–$F_p$ plane within a factor of □1.20 in both coordinates. For a few cases (~ 10) the region was not populated and so the coordinate ranges were enlarged 10% per iteration until at least one point was included in the region. The "best" representation was then computed by minimizing a statistic similar to absolute value norms discussed by Scargle (1981),

$$<d> = \{\Sigma\,|\tau_{model} - \tau_{real}|\,/\,(|\tau_{model}| + |\tau_{real}|)/2\}/N_{lags} \qquad (4)$$

searching for the minimum over the 5-parameter model space. A $\chi^2$ minimization is not appropriate in this situation, since the lag errors are not Gaussianly distributed over the $\tau_{lag}$–$F_p$ plane, but rather are strong functions of both peak flux and lag (and implicitly, a function of the disparate time profiles as well). The $<d>$ statistic instead minimizes fractional distance in model–real lag pairs (the $F_p$ coordinates being equal), where lag varies over ~ 3 orders of magnitude. The problem remaining in order to employ this statistic was, how to one-to-one associate a real lag and a model one? This problem was solved by decimation, finding the smallest distance in the $\tau_{lag}$–$\log_{10}(F_p)$ plane for a pair, with the decimation starting at the furthest distance from the approximate centroid, $\{\tau_{lag}, F_p\} = \{0.84, 1.2\}$. The decimation thus proceeded from the sparsely populated outer regions to the centroid, eliminating the closest model–real lag pair per step.

The resulting "best fit" set of model lags to the real lags in the $\tau_{lag}$–$F_p$ plane was

$$\tau_{model} (s) = 0.075 \{ 1 - R [1 - (\tau_{max}/\tau_{min})^{-0.15}] \}^{-1/1.15}$$
$$\text{with } \tau_{max} (s) = 0.125 (F_p/25.)^{-1.0} . \tag{5}$$

A typical realization of model lags is illustrated in Figure 9a, as gray-fill histograms on top of the measured lags (solid histograms) for the same six peak flux ranges shown in Figure 6. The general agreement appears satisfactory. Notably, the region near zero lag for $0.25 < F_p < 0.50$ is adequately modeled. Without inclusion of the $S_{obs}$ time-dilation factor, lags near zero tend to be overproduced in this lowest $F_p$ range. Note that as $\tau_{max}$ increases with lower $F_p$ – generating longer lags in the core – since fractional errors are large for $\tau_{lag} \sim 0$ s, the negative lags at low $F_p$ are fairly well reproduced by the model. In Figure 9b the same picture is illustrated except without errors for the model lags. Now the effect of the fixed value for $\alpha$ is apparent in the persistent peak near positive, short lags. In Figure 10 the model lags with errors are plotted in the $\tau_{lag}$–$F_p$ plane. Fidelity in representing the scatter plot for real lags (cf. Figure 2a) appears acceptable except for a handful of (real) long-lag outliers with $F_p > 3$ photons cm$^{-2}$ s$^{-1}$, which are not reproduced due to the fixed dependence of $\tau_{max}$ on $F_p$ – a minor defect for most purposes.

### 2.4 *Estimating Cosmological Corrections*

The last step is to correct the modeled lags (without noise) for time dilation and spectral redshift. The cosmological transformation can be realized by a procedure that generates a luminosity, $L_0$, from a lag-luminosity relation, where lag depends on z. The redshift is varied until the modeled peak flux agrees with that observed (our $F_p$, $\nu_1$–$\nu_2$ : 50–300 keV). The tutorial on distance measures in cosmology by Hogg (2000) is instructive, in which Eqs. (14–16, and 22) relate differential peak flux to luminosity distance, redshift, and differential luminosity in the source frame:

$$F_{p,model} = F_{p, \nu1-\nu2} = (1 + z) L_{(1 + z) \nu1-\nu2} / 4\pi D_L^2 . \tag{6}$$

An estimate of the luminosity in the "deredshifted" bandpass, $L_{(1 + z) \nu1-\nu2}$, can be gotten from the ratio of flux in the $(1 + z) \times (\nu_1-\nu_2)$ bandpass to a "bolometric" flux, multiplied by $L_0$. To calculate the flux ratio a simple model for the average GRB spectrum was assumed, a broken power law with fixed low and high energy spectral indices. The parameter values adopted are the approximate modes of the analysis by Preece et al. (2000) for the Band model (Band et al. 1993):

$$N_\gamma(E) = E^{-\alpha} = E^{-1} \qquad E < E_{join} = 230 \text{ keV}$$
$$= E^{-\beta} = E^{-2.25} \qquad E \geq E_{join}. \qquad (7)$$

The difference between a Band model and broken power-law is of no consequence in this treatment. In fact, the differences in results obtained for β = 2 or 2.25 are also negligible. Then the expression for the band-limited luminosity in the source frame is

$$L_{(1+z)\,\nu1-\nu2} = L_0 \int_{(1+z)\nu1}^{(1+z)\nu2} E\, N_\gamma(E)\, dE \Big/ \int_{E_{min}}^{E_{max}} E\, N_\gamma(E)\, dE \qquad (8)$$

where the limits on the γ-ray bolometric flux were chosen to be $E_{max}$ = 25 keV and $E_{min}$ = 20 MeV. While only ~ 20 burst spectra have been measured with the EGRET calorimeter, the average GRB spectrum is inferred to continue as a power law into the 10s of MeV regime (Dingus 2000). In principle, a full treatment should take into account spectral variation correlated with lag and peak flux. But since the variation in high energy power-law index noted above influenced the results negligibly, this approach with a fixed spectrum is probably a reasonable approximation.

The lag-luminosity relation needs to be augmented in this model to include lags longer than a few 100 ms, i.e., to include the long-lag bursts of apparently ultra-low luminosity. This other branch of the "GRB HR diagram" (NMB) is empirically constrained only by GRB 980425 and the absence of any bursts on the main branch with lags longer than ~ 300 ms. Hence, a partially arbitrary conjecture to satisfy this lack of constraint would have the hypothetical low-luminosity branch commence near 350 ms, or $L_{53} \approx 2.2 \times 10^{-2}$ (see also the similar, but theoretically based conjecture of Salmonson 2001), and thus

$$L_0 = L_{53} = 1.3 \times [\tau_{lag}/0.01 \text{ s}]^{-1.15} \qquad 0.003 < \tau_{lag} < 0.35$$
$$= 7.8 \times [\tau_{lag}/0.1 \text{ s}]^{-4.7} \qquad 0.35 < \tau_{lag}. \qquad (9)$$

The second defining point for the low-luminosity branch is GRB 980425, for which $L_{53}$ ~ 1.25 × $10^{47}$ ergs s$^{-1}$ and the measured $\tau_{lag}$ = 2.8 s (cf. NMB, where the lag for the whole profile was estimated visually as ~ 4.5 s).

The remaining ingredient is correction of the model lags for the extrinsic effects of time dilation and spectral redshift. The former correction factor is trivial, (1 + z). The correction for redshift of temporal structure from the fiducial 25–50 keV and 100–300 keV bands in the source frame, to lower energies in the detector frame, can be only approximately modeled using measurements of the brightest BATSE bursts' average temporal structure (Fenimore et al. 1995;

Norris et al. 1996). These bursts will tend to be at lower redshift than dimmer, short-lag bursts. At a redshift of $1 + z = 5$ (10), the 25–50 keV band is shifted to 5–30 keV (2.5–15 keV). No quantitative narrow-band measurements of the relative widths of temporal structure have been reported in the X-ray bands. Visual inspection by this author of BeppoSAX and older X-ray temporal profiles of GRBs appears to indicate that pulse structures broaden faster per logarithmic interval at these low energies than at BATSE energies. Here for expediency, the average pulse width of bright BATSE bursts is utilized to correct roughly for redshift, $W_{<pulse>} \propto E^{-0.33}$ (Norris et al. 1996), but with the exponent modified by a sigmoid function at high redshifts. Then the combined cosmological correction applied to $\tau_{model}$ of Eq. (5) is

$$S_{model} = (1 + z)^{-1} (1 + z)^{\kappa} \qquad (10)$$

with $\kappa = 1 - (2/3)\{2/[1 + \exp(z/z_{lim})]\}$. The expression reduces to $\kappa = 1/3$ at $z = 0$, and asymptotes to unity as z surpasses $z_{lim}$, where it nearly equals and counteracts time dilation (the effect of spectral redshift could be even larger: $\kappa$ could grow larger than 1). The lack of knowledge of $\kappa$ at low energies represents the largest source of uncertainty in the whole treatment. Hence the model is evaluated for two illustrative trial values of $z_{lim}$, 5 and 20.

The iteration procedure then proceeds as follows: Assume the cosmology $\{\Omega_M, \Omega_\Lambda\} = \{0.3, 0.7\}$ with $H_0 = 72$ km/s/Mpc (Freedman et al. 2001). Guess an initial redshift ($z_0$), compute the cosmology-corrected lag from Eq. (10), and integrate the luminosity distance. The source frame bandpass and lag-predicted luminosities defined by Eqs. (8) and (9) yield a model peak flux (Eq. [6]) in the detector frame, $F_{p,model}$, to be compared with $F_{p,obs}$. The next guess is generated from $z_{next} = z_{old} (F_{p,model}/F_{p,obs})^-$. The procedure converges (1% accuracy in $F_p$) within a few iterations yielding the same $z_{final}$ whether $z_0 = 1$ or 0.

Figures 11 through 13 illustrate the immediate results obtained for spectral lags, redshifts, and proper distances, respectively, for $z_{lim} = 5$ (solid histograms), and 20 (dashed histograms). Note that these modeled distributions are representations for the BATSE *sensitivity-limited* sample of long bursts. A common feature in these figures is the necessary convergence of the distributions for both values of $z_{lim}$ for the putative nearby sources of long-lag GRBs. Also, all the distributions peak near values corresponding to large redshifts. The cosmology-corrected lags (Figure 11) peak near 7 ms ($z_{lim} = 20$) and 13 ms ($z_{lim} = 5$), the majority occurring in either case at $\tau_{lag} < 50$ ms, implying that the observed luminosity distribution is peaked in the neighborhood of $L_{53} \sim 1$. Figure 12 shows the redshift distribution plotted as $(1 + z)$, thereby artificially emphasizing the nearby sources. For $z_{lim} = 5$ (20) the sources at cosmological distances are most numerous near $z \sim 10$ (20). This uncertainty in where the GRB rate-density peaks just affirms our lack of knowledge of the appropriate form of the spectral correction for high redshift. Figure 13 again emphasizes the preponderance of sources at large proper distance. But the other interesting feature is that ~ 90 sources are modeled to lie at distances < 100 Mpc. A more accurate estimate of their distances would require better definition of the steeper lag-luminosity branch defined in Eq. (9). These apparently nearby sources may have interesting ramifications for other astrophysics topics such as gravitational radiation from aspherically collapsing objects and ultra-high energy cosmic rays, and for the future Swift mission (see Discussion).

In Figure 14 the scatter plots of peak flux versus redshift for the two values of $z_{lim}$ have the expected appearance: The hard limit of $L_{cutoff}$ is manifest in the upper right boundary of each populated region. For $z_{lim} = 20$ the sources actually pile up at this boundary, reflecting the increase of redshift to accommodate short-lag, high-luminosity bursts, suggesting that with $z_{lim} = 20$ the $(1 + z)^\kappa$ factor in Eq. (10) ramps up too slowly. For $z_{lim} = 5$ (20) the median z increases from ~ 2 (3) for bright bursts to ~ 10 (20) for the dimmest bursts. While star formation rates are highly uncertain at these redshifts, investigators do not currently envision that star formation peaks at z > 20 (e.g., Madau, Della Valle, & Panagia 1998), and so a $z_{lim}$ as high as 20 in Eq. (5) is not favored. Below $F_p$ ~ 4 photons $cm^{-2}$ $s^{-1}$, the concentration of low-luminosity, long-lag bursts appears near z = 0. Besides these obviously expected features, there is a sparsely populated triangular region at low peak flux between the nearby low-luminosity GRBs and the high-luminosity cosmological GRBs. This sparsely populated zone (z < 2.3 near BATSE threshold) is consistent with a "seeing through" of the GRB population since lower redshift sources should continue to be detected at the same peak flux where higher redshift sources are represented.

It is interesting to compare the current distribution of 24 redshifts obtained from GRB afterglows or host galaxies, shown in Figure 15 (see the GCN alerts, maintained by Scott Barthelmy: http://gcn.gsfc.nasa.gov/gcn/gcn3_archive.html), with the $F_p$–z scatter plot in Figure 14b. The mode and median for determined redshifts continue to hold near z ≈ 1, with ~ 67% of the sample contained within 0.4 < z < 2; whereas for the model above $F_p$ ~ 10 photons $cm^{-2}$ $s^{-1}$ the median is z ~ 2.5. Approximately 120 bursts have been detected by BeppoSAX, RXTE, HETE-II, and/or the interplanetary network since the "afterglow era" began in February 1997. In the overwhelming majority of cases when X-ray observations were made, X-ray afterglows were detected, but for only ~ 40–50% of these were optical afterglows detected, perhaps half the time leading to host galaxy identifications with redshifts associated to the GRB (Kevin Hurley, private communications). Thus obscuration at the source or redshift of the spectrum have prevented optical afterglow detection, probably preferentially for higher redshift GRB sources.

## 3. COMPARISON OF MODEL AND THEORETICAL LUMINOSITY DISTRIBUTIONS

For the BATSE sensitivity-limited sample of long duration ($T_{90} > 2$ s) bursts, the luminosity distribution ($z_{lim} = 5$) derived from the model lags is illustrated in Figure 16. On the right side of the figure the seven cascading dotted/solid histograms are for the redshift-limited cuts: z < 30, 20, 10, 5, 3, 2, and 1. The plot is rendered in *equal log-spaced* intervals; hence the slopes of the two fitted power laws that are illustrated appear one unit flatter than for linear-spaced intervals. The distributions for z < 5, 10, 20, and 30 show increasing numbers of bursts at higher luminosities – the *observed* luminosity distribution peaks near $L_{53} \sim 1$. But this just reflects the deficit of detectable lower luminosity bursts at higher redshift. GRBs with $L_{53} < 10^{-4.6}$ (leftmost dotted histogram, z < 0.024, d < 100 Mpc) comprise the subsample with $\tau_{lag} < 2$ s. Recall that the model employs the two-branch lag-luminosity relation expressed in Eq. (9), which breaks at $\tau_{lag} = 0.35$ s, or $L_{53} = 2.2 \times 10^{-2}$. The number of long-lag bursts per logarithmic interval decreases as lag increases across the break point. The trend can be roughly described as a power-law, $dN/d\tau_{lag} \sim \tau_{lag}^{-1.8}$ over $0.01 < \tau_{lag}$ (s) < 3 (see Figure 11). Thus it is mostly the steeper power law of the low-luminosity branch which gives rise to lower frequency of GRBs per logarithmic interval for $L_{53} < 2 \times 10^{-2}$.

As noted, the model luminosity distribution reflects the sensitivity-limited BATSE sample. However, since we know something about GRB redshifts, a high-luminosity volume-limited regime can be rendered from this picture for particular cuts in luminosity and redshift. From inspection of Figure 14 the suggestion (section 2.3) was that BATSE has sampled through the GRB population for z < 2.3. Consider a source number density $\eta(L)dL = dN/dV$, where $\eta(L)$ may implicitly include a luminosity dependence for viewing angle, a profiled jet, etc. The number of sources actually detected (in Euclidean space) is then

$$N(L)dL = \int_0^{\rho} \eta(L)dL \, 4\pi r^2 \, dr \qquad (11)$$

where $\rho = \min\{R_0^3, [L/4\pi F_{p,thres}]^{3/2}\}$ and $F_{p,thres}$ is the peak flux at BATSE trigger threshold. Thus for $R_0 < [L/4\pi F_{p,thres}]^{1/2}$, a sample is volume limited, $N(L)dL \propto \eta(L)dL$, and we see the intrinsic luminosity distribution. (Any evolution with redshift is ignored in evaluating the average N(L); evolution may be a discriminant for GRB models, but it is not examined here). Beyond $z \sim 0.2$, the luminosity distance must be used since it significantly exceeds the Euclidean distance. From Eq. (7) the average photon energy is 135 keV, and so $F_{p,thres} \approx 5.4 \times 10^{-8}$ erg cm$^{-2}$ s$^{-1}$. Then for z < 2 and $L_{53} > 1.5 \times 10^{-2}$, $D_L(z=2) = 15.1$ Gpc, and $[L/4\pi F_{p,thres}]^{1/2} = 15.2$ Gpc, and so this constitutes a volume-limited regime. integration of Eq. (11) yields N(L)dL

η(L)dL. Over the volume-limited regime the distribution has a fitted power-law slope of −0.82 (solid straight line), so

$$dN_{vol}/dL \propto L^{-1.8}, \quad \text{and}$$
$$\eta(L) \propto L^{-1.8} \quad 10^{-1.6} < L_{53} < 10^{0.6}. \quad (12)$$

Whereas, for z < 1 and $L_{53} < 2.6 \times 10^{-3}$, $D_L(z=1) = 6.4$ Gpc and $[L/4\pi F_{p,thres}]^{1/2} = 6.3$, and this regime is sensitivity limited – the sources are progressively undersampled as luminosity decreases across this regime. Integration of Eq. (11) now yields $N(L)dL \propto \eta(L)L^{3/2}dL$. For the sensitivity-limited regime the fitted power-law slope is +0.015 (dashed straight line), so

$$dN_{sen}/dL \propto L^{-1}, \quad \text{and}$$
$$\eta(L) \propto L^{-5/2} \quad 10^{-6.4} < L_{53} < 10^{-2.6}. \quad (13)$$

(If the lag breakpoint for the two-branch lag-luminosity relation in Eq. [9] is increased from 0.35 s to 0.6 s, then $L \propto [\tau_{lag}/0.1\ s]^{-5.9}$, and the fitted power law slope becomes −0.015; hence the slope for the sensitivity-limited regime is not very dependent on the precise break point position.) Below $L_{53} \sim 10^{-6.4}$ the detected source frequency falls off, indicating either that such low-luminosity GRBs are below the BATSE trigger threshold, or perhaps that the intrinsic distribution cuts off. For reference, the burst with lowest determined redshift and luminosity, ~ $1.25 \times 10^{-6} L_{53}$ for GRB 980425 (z = 0.0085, d = 38 Mpc, $F_p = 0.9$ photons cm$^{-2}$ s$^{-1}$), would have been detected by BATSE at the sample threshold, $F_p \sim 0.25$ photons cm$^{-2}$ s$^{-1}$, at ~ 72 Mpc.

Recall the handful (~ 15) of real long-lag outliers ($F_{p,obs} > 3$ photons cm$^{-2}$ s$^{-1}$) not reproduced at comparable $F_{p,model}$ by the model lags (cf. Figure 2a and Figure 10) due to the fixed (versus fuzzy) dependence of $\tau_{max}$ on $F_p$. This represents a truly minor defect in the modeling among the ~ 210 bursts with $L_{53} < 10^{-2}$ over 5 decades in luminosity: According to their real positions in the $\tau_{lag}$–$F_p$ plane, about half would tend to fall in the underpopulated region $< 4 \times 10^{-7} L_{53}$, and the rest would be distributed fairly uniformly up to ~ $10^{-3} L_{53}$.

### 3.1 *Luminosity Distribution via Range of Viewing Angle, Constant Γ*

Eqs. (12) and (13) can be compared to expectations from purely relativistic kinematic proposals that viewing angle, $\theta_v$, off the jet axis is the governing factor for the observed luminosity distribution (see Salmonson 2000; 2001). For instance, in Salmonson & Galama (2001) the luminosity scales (neglecting redshift) as the perceived Doppler factor, $L \propto D \sim 2\Gamma / (1 + [(\theta_v - \theta_{jet})\Gamma]^2)$ for $\theta_v \ll 1$ and $\Gamma \gg 1$, i.e., near the jet cone, $\theta_{jet}$. But for $\theta_v$ large compared to $\theta_{jet}$ (~ few degrees), $L \sim D^3$. and the GRB appears as a low-luminosity source. In the latter case, for $\Gamma \sim$ few and constant, $L \sim [\theta_v\Gamma]^{-6} \propto \theta_v^{-6}$. The perceived source density distribution in the pure viewing angle scenario is $\eta(L)dL \propto \sin(\theta_v)d\theta_v$. Even for $\theta_v < 40°$, $\eta(L)dL \sim \theta_v d\theta_v$, and so $\eta(L)dL \sim L^{-4/3}dL$, and $N(L)dL \propto \eta(L)L^{3/2}dL \sim L^{+1/6}dL$. This slope, more than one unit steeper than Eq. (13), would result in a factor of $\sim 3000$ overproduction of GRBs at $10^{-6}L_{53}$ compared to $10^{-3}L_{53}$. This is consistent with the approximation made in the solid angle integration, given that $\theta_v$ varying over a factor of 10 (say, $\theta_{v,min} \sim 4° > \theta_{jet}$) would result in a dynamic range of $\sim 10^6$ in luminosity. Thus the extended flat distribution provides a good discriminant in the sensitivity-limited regime against the purely relativistic kinematic explanation for the low end of the modeled GRB luminosity distribution.

For the high-luminosity, volume-limited regime, $10^{-1.6} < L_{53} < 10^{0.6}$ with $z < 2$, the picture is less straightforward to interpret, since a significant fraction of bursts will still be observed within the jet cone – and if the beaming fraction varies appreciably (possibly correlated with $\Gamma$), we are in mixed territory (which may be the actual situation). Still, keeping with the purely relativistic kinematic interpretation and $\theta_{jet} <$ few degrees, since $\Gamma >> 1$, $L \sim [(\theta_v - \theta_{jet})\Gamma]^{-2} = [\theta_\Delta\Gamma]^{-2}$. Then via the same reasoning as above, $\eta(L)dL \sim L^{-2}dL$, not significantly different from Eq. (12). However, any realistic dependence of the Lorentz factor on angle with respect to the jet axis ($\Gamma \propto \theta_{jet}^{-\lambda}$, $\lambda > 0$) will render $\eta(L)$ steeper than $L^{-2}$. So again, the purely relativistic kinematic explanation is not favored to explain the distribution of highly luminous GRBs.

### 3.2 *Luminosity Distribution via Variable Beaming Fraction*

In the pure beaming fraction scenario, the luminosity and energy radiated into a differential solid angle are taken to be constant across the face of the jet cone, and negligible outside it. Also, the luminosity radiated into $4\pi$ steradians is found to be an approximate invariant (Frail et al. 2001), such that $L^{-1} \propto f_b \equiv \Omega_{jet}/4\pi$. To reproduce the modeled $\eta(L)$ dependences in Eqs. (12) and (13) requires distributions in opening angle, or jet-cone solid angle, $dN(\Omega_{jet})/d\Omega_{jet} \propto \Omega_{jet}^{-\lambda}$. It follows that $\lambda = 0.2$ for the volume-limited regime,

$$dN(\Omega_{jet})/d\Omega_{jet} \propto \Omega_{jet}^{-0.2}, \quad \theta < 20° \qquad (14a)$$

and $\lambda = -0.5$ for the sensitivity-limited regime,

$$dN(\Omega_{jet})/d\Omega_{jet} \propto \Omega_{jet}^{+0.5}, \quad \theta > 20°. \qquad (14b)$$

The break at $\theta = 20°$ ($f_b = 0.06$) is evaluated for $\tau_{lag} = 0.35$ in Eq. (1), corresponding to the break in the assumed two-branch lag-luminosity relation of Eq. (9). For the high-luminosity GRBs, Eq. (14a) predicts a slowly decreasing distribution for $\Omega_{jet}$. Thus, as the opening angle increases from the minimum to maximum values derived by Frail et al. (2001) and PK ($\theta_{jet} \sim$ 2–20 ), $dN(\Omega_{jet})/d\Omega_{jet}$ decreases by only a factor of ~ 0.6. For the low-luminosity GRBs, Eq. (14b) predicts that the number of bursts actually increases approximately linearly with $\theta_{jet}$ (since $\Omega_{jet}/4\pi \approx \_ \theta_{jet}^2$).

Now, employing the two branches of Eq. (1) that relate spectral lag and beaming fraction (and remembering that the relation for the long-lag branch is merely a toy model), the model lags of Figure 11 can be translated into numbers of bursts which would have been detected by BATSE if we had observed them all within their jet cones. This distribution, $N[f_b(\tau_{lag})] / f_b$, is plotted in Figure 17 for $z_{lim} = 5$. The implied number of highly luminous bursts ($f_b \_ 3 \times 10^{-3}$, $\tau_{lag} < 0.067$ s, $L > 10^{-0.8} L_{53}$) – most of them sufficiently luminous to be seen across the Universe – would need to be reproduced by a successful physical model. Only ~ 72% (1437/~2000 = $f_{use}$) of BATSE long bursts were analyzed in this study. The corrected rate is $N_{GRB} \_ 6.6 \times 10^5$ / $(9.2 \times f_{BATSE} f_{use})$ yr$^{-1}$, where the average live fraction for BATSE was $f_{BATSE} \sim 0.48$ (Hakkila et al. 2002). Therefore the rate of highly luminous bursts, irrespective of actual jet axis orientation, is estimated to be $N_{GRB} > 2 \times 10^5$ yr$^{-1}$.

### 3.3 *Luminosity Distribution via a Profiled Jet*

Models invoking a profiled jet with $\Gamma$, and therefore luminosity, decreasing off the jet axis have been developed recently by several authors. This dynamical mechanism can account for the anti-correlation between afterglow break times and luminosity, and possibly spectral evolution in the γ-ray phase (MacFadyen, Woosley, & Heger 2001; Salmonson & Galama 2001; Rossi, Lazziti, & Rees 2002; Zhang & Meszaros 2002). Consider $L_{jet} \propto \Theta_v^{-\gamma}$ with $\gamma > 0$. Then the solid angle dependence $\eta(L)dL \sim \Theta_v d\Theta_v$ yields $\eta(L)dL \sim \Theta_v^{\gamma+2} dL$, and $\eta(L)dL \sim L^{-1-2/\gamma} dL$. Agreement with the model distribution in the high-luminosity regime of Eq. (12) requires $-1-2/\gamma = -1.8$, or $\gamma = 5/2$:

$$L_{jet} \propto \Theta_v^{-5/2} . \qquad (15)$$

Within the profiled jet cone the observed luminosity is integrated over the observer viewable angle $\sim \Gamma^{-1}$, and so the luminosity dependence is an integration of dependences on $\Gamma(\Theta_v)$ as well as on $\Theta_v$, which are implicit in Eq. (15).

To estimate a lower limit on the number of highly luminous GRBs, a working model for the profiled jet must be assumed. For specificity, and to include the highest luminosity burst produced by the model lags, take $L(\Theta_v) = 3.6 L_{53} \times [\Theta_v/\Theta_0]^{-5/2}$ and two values for the minimum jet cone radius, $\Theta_0 = 1.5$  and 3  (0.026 and 0.052 radians, respectively), below which the luminosity remains constant (to avoid divergence) at $L = L(\Theta_0)$. Jets not oriented toward the observer increase the number of GRBs preceived to have $L(\Theta_v)$ by a factor $4\pi/[2\ 2\pi\Theta_v d\Theta_v]$, since the fraction of bursts with $L > L_{\Omega v}$ is $4\pi/[2\int^{\Omega v(L)}d\Omega_v]$ (the extra factor of 2 accounts for jets being two-sided). Figure 18a shows the resulting distribution of GRBs within the BATSE sensitivity reach for the cutoff $L > 10^{-1.5}L_{53}$, irrespective of jet axis orientation with respect the observer. If the source density were slightly steeper than expressed in Eq. (12), say $\eta(L) \propto L^{-2}$, then $\gamma = 2$. Figure 18b shows the distributions for $L_{jet} \propto \Theta_v^{-2}$ and the same two values of $\Theta_0$. GRBs with $L > 10^{-1.5}L_{53}$ are volume-limited within $z < 2$, while the most highly luminous GRBs, up to $L_{53} \sim 1$, are observable even to $z \sim 20$. To compare with the rate calculated above for beaming fraction with $L > 10^{-0.8}L_{53}$, for $L_{jet} \propto \Theta_v^{-2.5}$ the total numbers are $2.7 \times 10^6$ ($7 \times 10^5$) for $\Theta_0 = 1.5°$ (3°). For $L_{jet} \propto \Theta_v^{-2}$ the corresponding totals are $2.1 \times 10^6$ ($5.4 \times 10^5$). Then the rates are $R_{GRB} > N_{tot} / (9.2 \times f_{BATSE}f_{use})$ yr$^{-1} \approx 0.66$–$0.85 \times 10^6$ yr$^{-1}$ ($\Theta_0=1.5°$) and $1.7$–$2.2 \times 10^5$ yr$^{-1}$ ($\Theta_0=3°$). The lower limits on rates for the profiled jet and beaming fraction scenarios are compared with estimates of the SN Ib/c rate in the Discussion.

For the sensitivity-limited regime, Eq. (13) implies $L_{jet} \propto \Theta_v^{-4/3}$, a flatter dependence than for the highly luminous GRBs. Assuming that $\Gamma \sim$ few for the long-lag GRBs, the observable $\gamma$-ray cone ($\theta_\gamma \sim \Gamma^{-1}$) can be substantial fraction of $2\pi$. Hence for the low-luminosity GRBs, a large fraction of jets with axes not directly oriented toward the observer are still observed.

## 4. DISCUSSION

In the pre-afterglow GRB era it was often remarked that burst temporal-spectral behavior was chaotic – unpredictable – zoo-like, and discussions were held at workshops to attempt useful classification schemes. But prior to BATSE, such classifications were uninterpretatable because GRBs *were* largely unpredictable – on their duration timescales. However, the sensitivity of the BATSE experiment has allowed us to see and map an important organizing principle: pulse spectral evolution. Within a given burst the individual pulses tend to have comparable widths and spectral lag. But in the whole burst population, pulse widths and lags vary over a large dynamic range, $> 10^3$. This variation appears to be inversely related to luminosity in long ($T_{90} > 2$ s) BATSE bursts. (A second organizing principle in $\gamma$-ray time profiles, reported by Ramirez-Ruiz & Merloni [2001], is yet to be completely exploited: Time intervals near background level, between major emission episodes, are well correlated with the time interval of the following episode, strongly suggesting that a metastable accretion mechanism is fueling the burst process. Almost certainly, this correlation eliminates the external shock hypothesis as the energy release mechanism for the $\gamma$-ray phase [E. Ramirez-Ruiz, private communication].)

The wide dynamic range in perceived GRB luminosity has been revealed by afterglow observations which yield associated redshifts. And now the afterglows (or their absence) also manifest variable behavior – varying decay timescales and power-law break times. Many investigators have proposed unifying models which can explain the break times, dynamic ranges in luminosity and γ-ray energy, and in some cases the γ-ray pulse behavior (Frail et al. 2001; Panaitescu & Kumar 2001; Wang & Wheeler 1998; Höflich, Wheeler, & Wang 1999; Nakamura 1999; Woosley, Eastman, & Schmidt 1999; Salmonson 2001; Ioka & Nakamura 2001; Salmonson & Galama 2001; Rossi, Lazziti, & Rees 2002; Zhang & Meszaros 2002). The economy of each of the three model classes – variable beaming fraction, pure relativistic kinematics (viewing angle), and profiled jets with graded Lorentz factor – is difficult to evaluate: each affords a reasonable explanation of the dynamic ranges in GRB energy and luminosity, such that the actual energy release into 4π steradians is approximately constant, at least for GRBs at cosmological distances. In fact, reality may be a combination of the proposed effects. However, assuming a lag-luminosity relation, one can derive distributions for the large BATSE long-burst sample which constrain the three pure mechanisms described above. Here, I have analyzed spectral lags for 1437 long ($T_{90}$ < 2 s) BATSE bursts to near the trigger threshold, and modeled the lags to yield a noise-free representation. The model lags with cosmological corrections were used to compute distributions for redshift, proper distance, and luminosity. The main results, which may have further implications for GRB populations, are summarized:

A GRB subsample is identified – those with few, wide pulses, lags of a few tenths to several seconds, and soft spectra. The proportion of these long-lag bursts (Figure 3) increases from negligible among bright BATSE bursts to ~ 50% at trigger threshold. Their integral size–frequency distribution (Figure 4) follows a –3/2 power-law over 1_ decades in $F_p$. Bursts with very long lags, ~ 1–2 < $\tau_{lag}$ (s) < 10, show a tendency to concentrate near the Supergalactic Plane (Figure 5) with a quadrupole moment of ~ –0.10 ⎕ 0.04 (Table 1). GRB 980425 (SN 1998bw) is a member of this subsample of ~ 90 bursts with estimated distances < 100 Mpc (Figure 13) and ultra-low luminosities (Figure 16).

In two luminosity regimes, the model lags yield approximate power-law scaling relations, Eqs. (12) and (13), for their respective distributions. These empirical relations are compared with expectations for jet scenarios which invoke variable beaming fraction with approximately constant total energy, relativistic kinematics (pure viewing angle effect), or profiled jets with variable luminosity (i.e., variable Lorentz factor) off the jet axis. In the high-luminosity volume-limited regime, $10^{-1.6}$ < $L_{53}$ < $10^{0.6}$ with z < 2, the model distribution is $dN_{vol}/dL \sim L^{-1.8}$. For the low-luminosity GRBs in the sensitivity-limited regime $10^{-6.4}$ < $L_{53}$ < $10^{-2.6}$, the model distribution scales as $dN_{sen}/dL \sim L^{-1}$.

Both dependences can be reproduced with profiled jets, where $L_{jet} \propto \Theta_v^{-5/2}$ for the high-luminosity bursts, and $L_{jet} \propto \Theta_v^{-4/3}$ in the low-luminosity regime. The beaming fraction scenario can also fit the high-luminosity bursts with a fairly flat distribution for the jet-cone solid angle, $dN(\Omega_{jet})/d\Omega_{jet} \propto \Omega_{jet}^{-0.2}$. For the low-luminosity bursts a distribution that actually increases approximately linearly with $\theta_{jet}$ ($\Omega_{jet}/4\pi \approx \_ \theta_{jet}^2$) is required, $dN(\Omega_{jet})/d\Omega_{jet} \propto \Omega_{jet}^{+0.5}$. The pure viewing angle scenario produces a distribution which is too steep for the low-luminosity GRBs, $dN_{sen}(L)/dL \propto L^{+1/6}$. For the high-luminosity GRBs, $dN_{vol}/dL \propto L^{-2}$, only slightly steeper than the model distribution ($\propto L^{-1.8}$). However, this would leave no room for inclusion of a profiled Lorentz factor, expected from modeling collapsars (MacFadyen, Woosley, & Heger 2001), which would steepen the distribution.

The lag model of Eq. (4) and iteration procedure of section 2.4 for estimating redshift could be used in other programs to obtain results, e.g. for a volume-limited treatment of the GRB universe. Currently the largest uncertainty in any such treatment is our limited knowledge of spectral redshift corrections for high z bursts. In fact, in Eq. (9) $z_{lim}$ could be lower than 5, and κ could be larger than unity for large redshifts. Therefore, it is prudent to view the results for $z_{lim} = 5$ as much more useful for prediction compared to $z_{lim} = 20$ where the extrapolation range is large. However, the luminosity distribution for high-luminosity bursts (Eq. [12]) should be unaffected by these uncertainties since it was constructed for burst sources with z < 2 and $L_{53} > 10^{-1.5}$, a volume-limited regime for which the estimated redshift correction is reasonable. The modeled redshift distribution for GRBs peaks near z ~ 10, a result which has a large associated error, due to the uncertainty in correction for the effect on time profiles of redshifted GRB spectra beyond z ~ few.

### 4.1 *GRBs, SNe, Gravitational Waves, and UHECRs*

The GRB-SN connection began with GRB 980425 / SN 1998bw. The large number of long-lag bursts – like GRB 980425 in several respects – that are found mostly at low BATSE peak flux, suggest some requirements on SN Ib/c and collapsar models (MacFadyen & Woosley 1999). Many of these long-lag bursts are implied by the two-branch lag-luminosity relation to be so close (within 10s of Mpc) that we should expect to identify frequently the responsible collapsar/hypernova – regardless of what SN type it turns out to be – especially when alerted with Swift localizations (Barthelmy 2001). If these are SN Ib/c, and their predicted low Lorentz factors (Γ ~ 2–5: Kulkarni et al. 1998; Woosley & MacFadyen 1999; Salmonson 2001) are at least partially responsible for the ultra-low luminosities, then approximately all SN Ib/c are needed to account for long-lag bursts: Bloom et al. (1998) estimate ~ 0.3 SN Ib/c day$^{-1}$ within 100 $h_{65}^{-1}$ Mpc. Over the ~ 9.2 years of CGRO operations, the BATSE sample used here would have had exposure to ~ $10^3 \times f_{BATSE} f_{use}$ ~ 350 SN Ib/c within the same volume. There are 372 bursts in region 3 of Figure 2 with lags > 0.35 s, and 155 (72) bursts have lags > 1 (2) s. The sources of approximately 90 of these bursts are predicted (Figure 13) to lie within 100 Mpc. For $\Gamma$ ~ few, the cone of observable γ-ray emission is large, and so in combination with the flatter predicted dependence for a profiled jet ($L_{jet} \Box \Theta_v^{-4/3}$) than for the high-luminosity GRBs, the observer would be within the γ-ray emission cone of most low-luminosity GRBs. Thus, a large fraction, at least 25%, of SN Ib/c would participate in making long-lag GRBs, or else some unobserved nearby source type would be required. Notice that in a whole-Universe, volume-limited sample, the ultra-low luminosity bursts would therefore constitute the major variety of GRB (see discussion in MacFadyen, Woosley, & Heger 2001).

Confirmation that the sources of long-lag GRBs are relatively nearby could come from several directions. First, untriggered BATSE bursts – most with peak fluxes lying below the sample threshold of this study (0.25 photons cm$^{-2}$ s$^{-1}$) – should be predominantly long-lag bursts (see Figure 3). However, the localizations for these dim events have large error regions, and their idenfication as bona fide GRBs becomes more probabilistic with decreasing peak flux. Another possibility is that, since the matter distribution within 100 Mpc is not planar, cross-correlation of the positions of nearby galaxies (Hudson 1993) and long-lag GRBs may yield a higher confidence result than the simple quadrupole test. Also, the light curves of unusual SNe for which more than one detection exist may be extrapolated to the SN onset; however, this approach may be least likely to yield confirmation since most SNe within 100 Mpc during the CGRO era went undetected (and still do), and such statistical treatments usually require ad hoc assumptions. The highest guaranteed approach will be real-time alerts provided by Swift, allowing detection of SNe if they are indeed associated with long-lag GRBs.

Now consider the possible association of GRBs at cosmological distances with SNe. From van den Bergh & Tammann (1991) and Cappellaro et al. (1999) one may roughly estimate that ~ 10% of all local SNe are type Ib (or Ic). Whereas at high redshifts perhaps 20% of SNe are type Ib/c since these core collapse events were more frequent in the early Universe (see Figure 2 of Madau et al. 1998). Assuming ~ 1 SN s$^{-1}$ in the observable Universe for all types, then ~ 6 × 10$^7$ SN Ib/c occurred (at our epoch) during the CGRO mission. The pure beaming fraction scenario (Figure 17, section 3.2) predicts a rate > 2 × 10$^5$ yr$^{-1}$ for highly luminous GRBs (L > 10$^{-0.8}$ L$_{53}$), or 1.8 × 10$^6$ GRBs during CGRO. Thus at least 1/30 of the SN Ib/c population at high redshifts – presumably the highly unusual and energetic ones – would be required to produce the most luminous GRBs. However, if the volume-limited luminosity distribution continues to scale as expressed in Eq. (12) beyond z = 2, then the total rate of GRBs with 10$^{-1.6}$ < L$_{53}$ < 10$^{1.6}$, is ~ 4–5 times larger, or ~ 10$^6$ yr$^{-1}$. Then ~ 1/7 of the SN Ib/c population would be required to produce the cosmologically distant GRBs in the pure beaming fraction scenario. Similarly, for the profiled jet scenario, the total GRB rates would be $\sim 3 \times 10^6$ yr$^{-1}$ and $\sim 10^6$ yr$^{-1}$ for minimum jet cone radii of $\Theta_0=1.5$ and $\Theta_0=3$, respectively (Figure 18, section 3.3). Again, a large fraction of the SN Ib/c population would have to participate in making highly luminous GRBs in the profiled jet scenario.

Results similar to those found here for the GRB redshift distribution have been inferred by Fenimore & Ramirez-Ruiz (1999). In addition to the GRB rate-density peaking at large redshift ($z \sim 10$), Lloyd-Ronning, Fryer, & Ramirez-Ruiz (2001) find that the GRB luminosity distribution scales as $(1 + z)^{1.4\pm 0.2}$. They conclude that the star formation rate continues to increase to redshifts at least as high as $z \sim 10$, in order to provide a sufficiently high progenitor rate for GRBs. Inferences on star formation in the early Universe based on GRB analyses may be compared with quasar observations in which the Gunn-Peterson trough is detected – evidence of the epoch of reionization. Djorgovski et al. (2001) and Becker et al. (2001) conclude that the reionization era was complete near $z \sim 5$–$6$. Simulations suggest that the reionization era spanned $z \sim 15$ to $z \sim 5$, roughly consistent with GRBs appearing at redshifts of at least $z \sim 10$ (Gnedin 2000; Ciardi et al. 2000; Umemura, Nakamoto, & Susa 2001).

With the optimization of LIGO II, S/N levels of order few $\times 10^{-24}$ Hz$^{-\frac{1}{2}}$ should be detectable, depending upon particulars of instrument enhancement (Fritchel 2001). For some tuned configurations the LIGO II sensitivity peaks in the $\sim 100$–$500$ Hz range. Modeling of axisymmetric core collapse by Zwerger & Müller (1997) using two-dimensional Newtonian hydrodynamic code predicts peak strains of order $10^{-23}$ in the 100–800 Hz band for distances of 10 Mpc. These expectations are reinforced by general relativistic modeling performed by Dimmelmeier, Font, & Müller (2001) which agree with amplitudes from Newtonian models within $\sim 30\%$. Also, Fryer, Holz, & Hughes (2001) expect strains from bar modes of few $\times 10^{-23}$ to $\times 10^{-22}$ at 10 Mpc. For sources at 100 Mpc, the strains would be $10^{-24}$–$10^{-23}$. For a signal persisting in this band for a few cycles, the LIGO II sensitivity translates into a strain sensitivity of $\sim 3 \times 10^{-24} (300/\text{few})^{\frac{1}{2}} \sim 3 \times 10^{-23}$ for a S/N ratio of $\sim 1:1$. Thus a small fraction of the $\sim 90/(9.2 f_{use} f_{BATSE}) \sim 30$ long-lag bursts yr$^{-1}$ (over several years) within 100 Mpc expected from the BATSE sample analysis may be detectable by LIGO II, at least if the signals are co-added (see Finn 2001). Since the long-lag GRB sources appear to be nearby, their signals would not be time-dilated beyond LIGO's sensitivity range. Detection will require computation of accurate chirp templates for core collapse events, and knowledge of the GRB occurrence time and location (an argument for maintaining all-sky monitoring of GRBs).

The long-lag bursts might be sources of ultra-high energy cosmic rays (UHECRs). Nagano & Watson (2000) provide an complete summary of the field. Proposals for GRBs at cosmological distances ($\Gamma \sim 10^2$–$10^3$) as the source of UHECRs have been made by Vietri (1995) and Waxman (1995). Vietri shows that the cosmic ray energy flux $> 3 \times 10^{18}$ eV expected from cosmological GRBs is comparable to that observed, $5 \times 10^{-13}$ ergs cm$^{-2}$ s$^{-1}$ sr$^{-1}$. However, Stecker (2000) and Scully & Stecker (2002) argue that energy losses suffered by GRBs at cosmological distances would produce too sharp a cutoff in the UHECR spectrum, and would fall short in producing the flux by a factor of ~ 100–1000. The latest results from AGASA indicate a pronounced flattening of the spectrum $> 10^{19}$ eV compared to results from previous experiments (Watson 2001). Thus the salient question is, how does the UHECR flux from ~ 30 GRB yr$^{-1}$ within 100 Mpc and $\Gamma$ ~ few compare with that for 900 GRB yr$^{-1}$ Universe$^{-1}$ beamed at us with $\Gamma \sim 10^2$–$10^3$, and could the closer GRB sources provide the enhancement seen by AGASA $> 10^{19}$ eV? There is the perennial problem of (an)isotropy. The AGASA results possibly indicate point origins, implying nearby sources, but the interpretation is still open. For energies $> 10^{20}$ eV, source distances < 100 Mpc, and magnetic field scale lengths $\lambda \sim 10$ Mpc, the arrival directions are scrambled if B $> 10^{-9}$ G, a typical extragalactic field strength believed by many investigators (see Kronberg 1994, on extragalactic magnetic fields). Farrar & Piran (2000) argue for a ~ 0.1 μG extragalactic field, corresponding approximately to equipartition between the magnetic, gravitational, and thermal energies of the Virgo supercluster for tangled fields. They also favor magnetic sheet topologies rather than two-dimensional random walk, and so a shorter distance results in the same deflection (~ 1 μG would result in a 10 deflection for $\lambda \sim 100$ Mpc even at E $> 6 \times 10^{21}$ eV). Observations will eventually reveal (an)isotropy, but modeling is called for to constrain the possibilities for nearby GRBs.

It is interesting that no similar $\tau_{lag}$–$\tau_{break}$–$\theta_{jet}$ relationship has been established for the other major astrophysical jet phenomenon, active galactic nuclei (G. Madejski, private communication). If a such relationship exists in AGN, its non-detection may be attributable to insufficient density of X-ray/γ-ray observations in time and energy. For GRBs spectral lags can be determined on a relevant (and rapid) timescale that does not required scheduled, extended observations. Of course, to be more useful, this most fascinating phenomenon of GRBs will need to be observed simultaneously over a broader γ-ray energy range. Future planned and proposed missions such as Swift, GLAST, and EXIST should provide very useful observations of GRB temporal behavior. Swift (Barthelmy 2001), with its improved sensitivity over BATSE and lower energy coverage, should give a larger yield of ultra-low luminosity GRBs with immediate alerts, allowing searches for the corresponding supernovae.

For many useful hints, suggestions, and encouragement I wish to thank: Jerry Bonnell, Jordan Camp, Joan Centrella, Thomas Cline, Dale Frail, Neil Gehrels, Kevin Hurley, John Krizmanic, Pawan Kumer, Greg Madejski, Miguel Morales, Robert Nemiroff, Jay Salmonson, and Jeff Scargle. Special thanks to David Band for a careful reading of the manuscript.

TABLE 1

Quadrupole Moment in Supergalactic Coordinates for Long-Lag Bursts

| $\tau_{lag} >$ | $N_{GRBs}$ | Q | $\varepsilon_Q$ | $N\sigma$ |
|---|---|---|---|---|
| 0.50 | 298 | -0.025 | 0.019 | 1.27 |
| 1.00 | 155 | -0.047 | 0.026 | 1.82 |
| 1.50 | 96 | -0.068 | 0.032 | 2.15 |
| 2.00 | 72 | -0.099 | 0.036 | 2.74 |
| 2.50 | 54 | -0.109 | 0.042 | 2.63 |
| 3.00 | 41 | -0.105 | 0.047 | 2.22 |
| 3.50 | 27 | -0.069 | 0.058 | 1.18 |

Figure captions

Fig. 1 – Inferred afterglow beaming fraction vs. gamma-ray spectral lag for eight bursts with associated redshifts and BATSE data. The lag is corrected for time dilation, and approximately corrected for redshift of the spectral energy distribution (see section 2.4, Eq. [10]). Solid line, described by Eq. (1a), was fitted only to the six bursts with $0.835 < z < 1.619$; dashed line is toy model of Eq. (1b) for long-lag regime. Diamonds for beaming fractions derived by Frail et al. (2001); squares for those derived by Panaitescu & Kumar (2001).

Fig. 2 – (a) BATSE peak flux ($F_p$, 50–300 keV) vs. CCF spectral lag ($\tau_{lag}$, 25–50 keV to 50–300 keV) for 1437 long-duration ($T_{90} > 2$ s) bursts. Dashed horizontal line indicates approximate BATSE trigger threshold. Solid lines divide the $F_p$–$\tau_{lag}$ plane into three regions: bright short-lag bursts (region 1), dim short-lag bursts (region 2), and long-lag bursts (region 3). The point for GRB 980425 is circled. (b) The $\tau_{lag}$ coordinate is magnified, showing only lags in the range $-0.2 < \tau_{lag}$ (s) $< 1.0$. Negative lags are completely consistent with larger errors at lower S/N.

Fig. 3 – Ratio of numbers of bursts in regions 1+2 to region 3, $N_{long-lag} / N_{short-lag}$, vs. peak flux binned in dyadic steps. Annotation shows that, while the sizes of both subsamples decrease near BATSE threshold, the increase in the relative frequency of long-lag bursts is significant.

Fig. 4 – The integral peak flux–frequency distributions (Log $N[>F_p]$ – Log $F_p$) for the three subsample regions delineated in Figure 2. Notably, the long-lag bursts follow a –3/2 power-law over a factor of ~ 40 in peak flux.

Fig. 5 – Sky distribution in the Supergalactic coordinate system for very long-lag ($\tau_{lag} > 2$ s) bursts (diamonds), suggestive of a tendency to concentrate toward the Supergalactic Plane. Circle marks center of Virgo Cluster. Boxed diamond indicates position of GRB 980425; empty diamond is position of GRB 971208 (~ 20–30 s lag). Table 1 lists significances of quadrupole moment for different cuts in $\tau_{lag}$. $Q = -0.099 \pm 0.036$ for illustrated distribution.

Fig. 6 – Spectral lag distributions for peak flux ranges binned in dyadic steps.

Fig. 7 – Similar to Figure 6, but with $\tau_{lag}$ magnified to show the core region for short-lag bursts. Overplotted histograms with gray fill are for brightest subsample ($100 > F_p$ [photons cm$^{-2}$ s$^{-1}$] $>$ 8) S/N equalized to three lowest $F_p$ ranges, demonstrating that lower S/N level alone does not account for the broadening of the $\tau_{lag}$ distribution at lower peak fluxes.

Fig. 8 – Observed time-dilation factor vs. peak flux for $T_{50}$ durations of those bursts in regions 1+2, with fitted quadratic model.

Fig. 9 – (a) The spectral lag distributions of Figure 6, overplotted (gray fill) with model lags including empirical errors. (b) Overplotted with the same model lags but without empirical errors.

Fig. 10 – Similar to Figure 2a: scatter plot of $F_p$ vs. model $\tau_{lag}$ including empirical errors.

Fig. 11 – Distribution of model lag with cosmological corrections for time dilation and spectral redshift for the two parameterizations of $\kappa$ in Eq. (10), $z_{lim}$ = 5 (solid) and 20 (dashed).

Fig. 12 – Distribution of redshift, derived by varying z to match model peak flux to observed BATSE peak flux, assuming the two-branch relationship between $\tau_{lag}$ and luminosity of Eq. (9); $z_{lim}$ = 5 (solid) and 20 (dashed). Abscissa, expressed as (1 + z), overemphasizes frequency of long-lag bursts, placing them in one bin.

Fig. 13 – Distribution of proper distances associated with redshifts of Figure 12; $z_{lim}$ = 5 (solid) and 20 (dashed). Approximately 90 modeled burst sources lie at distances less than 100 Mpc.

Fig. 14 – Scatter plot of peak flux versus redshift (1 + z). (a) $z_{lim}$ = 5, (b) $z_{lim}$ = 100. The hard limit of $L_{cutoff}$ is manifest in the upper right bound to the populated region. For $z_{lim}$ = 20 sources actually pile up at this boundary, reflecting increasing redshift to accommodate short-lag, high luminosity bursts (see text). Below $F_p \sim$ 4 photons cm$^{-2}$ s$^{-1}$, the concentration of low-luminosity, long-lag bursts appears near z = 0.

Fig. 15 – Distribution of 24 redshifts determined to be associated with GRB afterglows or the host galaxy (see Barthelmy, GCN alerts: http://gcn.gsfc.nasa.gov/gcn/gcn3_archive.html). The mode and median continue to hold near z $\approx$ 1, with ~ 2/3 of this sample within 0.4 < z < 2. Compare to modeled redshifts in Figure 14b for bright bursts, $F_p$ > 10 photons cm$^{-2}$ s$^{-1}$.

Fig. 16 – Luminosity distribution in *equal log-spaced* intervals predicted by the two-branch lag-luminosity relation expressed in Eq. (9) for the modeled lags ($z_{lim} = 5$) of the BATSE sample. The seven cascading dotted/solid histograms on right are for redshift-limited cuts: z < 30, 20, 10, 5, 3, 2, and 1. In the regime z < 2 over the two decades $10^{-1.5} < L_{53} < 10^{+0.5}$ the distribution is volume-limited, with fitted power-law slope of –0.82 (solid straight line), so $dN_{vol}/dL \propto L^{-1.8}$. For z < 1 within the range $10^{-6.4} < L_{53} < 10^{-2}$ the distribution is sensitivity limited, with power-law slope of +0.01 (dashed straight line), so $dN_{sen}/dL \propto L^{-1}$. Sources with $L_{53} < 10^{-4.6}$ (leftmost dotted histogram, z < 0.024, d < 100 Mpc) comprise the subsample with $\tau_{lag} < 2$ s.

Fig. 17 – Beaming fraction distribution for modeled lags of Figure 11 ($z_{lim} = 5$) divided by $f_b$, yielding distribution of all sources within BATSE reach, irrespective of beaming direction. This is still a sensitivity-limited distribution, with the volume sampled decreasing as $f_b$ increases. Implied rate of highly luminous bursts ($f_b \sim 3 \times 10^{-3}$, $L > 10^{-1.5} L_{53}$) is $N_{GRB} \sim 2 \times 10^5$ yr$^{-1}$.

Fig. 18 – For profiled jet scenario: GRB sources within BATSE sensitivity reach irrespective of jet axis orientation with respect to observer. Luminosity dependence is $L(\Theta_v) = 3.6 L_{53} \times [\Theta_v/\Theta_0]^{-\gamma}$ with cutoff $L > 10^{-1.5} L_{53}$, for two values of the minimum jet cone radius, $\Theta_0 = 1.5$ (solid histograms) and 3 (dashed histograms), corresponding to 0.026 and 0.052 radians, respectively. In (a) $\gamma = 2.5$, in (b) $\gamma = 2.0$. Accounting for sample and BATSE exposure results in 0.66–0.85 × 10$^6$ yr$^{-1}$ ($\Theta_0=1.5$) and 1.7–2.2 × 10$^5$ yr$^{-1}$ ($\Theta_0=3$) for the most highly luminous GRBs ($L > 10^{-1.5} L_{53}$).